\documentclass[fleqn,usenatbib]{mnras}

\usepackage{newtxtext,newtxmath}

\usepackage[T1]{fontenc}
\usepackage{ae,aecompl}

\usepackage{graphicx}	
\usepackage{amsmath}	
\usepackage{amssymb}	
\usepackage{etoolbox}

\makeatletter 
\patchcmd\@combinedblfloats{\box\@outputbox}{\unvbox\@outputbox}{}{%
   \errmessage{\noexpand\@combinedblfloats could not be patched}%
}%
 \makeatother

\title[The 21-cm signal around the first star]{Stellar mass dependence of the 21-cm signal around the first star and its impact on the global signal}

\author[T. Tanaka et al.]{
Toshiyuki Tanaka,$^{1}$\thanks{E-mail: tanaka.toshiyuki@d.mbox.nagoya-u.ac.jp}
Kenji Hasegawa,$^{1}$
Hidenobu Yajima,$^{2,~3}$
\newauthor
~Masato I.N. Kobayashi$^{1,~4}$
and Naoshi Sugiyama$^{1,~5,~6}$
\\
$^{1}$Division of Particle and Astrophysical Science, Graduate School of Science, Nagoya University, Chikusa, Nagoya 464-8602, Japan\\
$^{2}$Frontier Research Institute for Interdisciplinary Sciences, Tohoku University, Sendai 980-8578, Japan\\
$^{3}$Astronomical Institute, Tohoku University, Sendai 980-8578, Japan\\
$^{4}$Department of Earth and Space Science, Graduate School of Science, Osaka University, Toyonaka, Osaka 560-0043, Japan\\
$^{5}$Kobayashi-Maskawa Institute for the Origin of Particles and the Universe, Nagoya University, Chikusa, Nagoya, 464-8602, Japan\\
$^{6}$Kavli Institute for the Physics and Mathematics of the Universe (Kavli IPMU),
The University of Tokyo, Chiba 277-8582, Japan
}

\date{Accepted XXX. Received YYY; in original form ZZZ}

\pubyear{2018}

\begin{document}
\label{firstpage}
\pagerange{\pageref{firstpage}--\pageref{lastpage}}
\maketitle

\begin{abstract}
The 21-cm signal in the vicinity of the first stars is expected to reflect properties of the first stars. 
In this study we pay special attention to tracing the time evolution of the ionizing photons' escape fraction, which affects the distribution of neutral hydrogen, by performing radiation hydrodynamics (RHD) simulations resolving dense gas in a halo. 
We find that the radial profile of 21-cm differential brightness temperature is quite sensitive to the stellar and halo masses, which reflects the time evolution of the escape fraction.
In the case of a less massive star, ionizing photons hardly escape from its host halo due to the absorption by dense halo gas, thus an deep 21-cm absorption feature at just outside the halo lasts a long time. Whereas photons from a massive star well working to heat the ambient intergalactic medium turn out to cause a spatially extended 21-cm emission signature. 
Although individual signals are found to be undetectable with the Square Kilometre Array, our analysis using the results from the RHD simulations indicates that the properties of the first stars are imprinted on the 21-cm global signal: its amplitude depends not only on the cosmic star formation rate density, but also on the typical mass of the first stars due to the stellar-mass-dependent heating rate. Thus, we suggest that the initial mass function of the first stars is an essential factor in understanding the global signal.
\end{abstract}

\begin{keywords}
(cosmology:) dark ages, reionization, first stars -- radiative transfer -- hydrodynamics -- (galaxies:) intergalactic medium -- (ISM:) H II regions -- radio lines: general
\end{keywords}


\section{Introduction} \label{intro}
The first stars are believed to have formed at redshift $z = 20-30$ in dark matter mini-haloes with masses of $\sim 10^{5-6} \mathrm{M}_{\sun}$ \citep[e.g.,][]{Tegmark1997, Nishi1999, Abel2002, Bromm2002, Yoshida2003, Yoshida2006, Gao2007, O'Shea2007, Yoshida2008}, and to have played crucial roles on the subsequent structure formation and the thermal history of the Universe. 
For example, they are likely main ionizing sources at the beginning of cosmic reionization \citep{Alvarez2006, Johnson2007}. 
In addition, heavy elements provided by the first stars enhance the subsequent star formation activities \citep{Wise2012, Karlsson2013}. 
The first stars are also expected to be seeds of the super massive black holes observed at high redshifts 
 $z > 6$ \citep{Venemans2013, Banados2014, Wu2015}.

These impacts strongly depend on the mass spectrum of the first stars, thus clarifying the formation of the first stars is a crucial issue in the modern cosmology and astrophysics. 
It has long been known that the thermal evolution of the primordial gas is different from that at the present-day due to the lack of heavy elements \citep{Palla1983}. 
The typical temperature in a primordial gas cloud is higher than that in metal-rich gas clouds because of inefficient cooling by hydrogen molecules alone. 
This causes the high mass accretion rate onto a primordial protostar from a primordial gas cloud.
Recent computational power starts to allow us to consider the detailed physical processes and to evaluate the mass accretion onto the first protostar \citep{Hosokawa2011, Greif2012,Susa2013,Stacy2016}. 
However there is still inconsistency on the typical mass of the first stars even among recent state-of-the-arts simulations; 
For instance, \citet{Susa2014} predicted the mass function with a peak at $\sim100~\rm M_{\sun}$, while \citet{Hirano2015} showed a more wider range extended to more massive stars. 

As for observations, the direct detection of the stellar radiation from the first stars is desirable but is still difficult even with the forthcoming instruments, such as the European Extremely Large Telescope, the James Webb Space Telescope, and the Thirty Meter Telescope. 
An alternative way to explore the evidence for the first stars is to observe the hyperfine structure line of neutral hydrogen, so-called 21-cm line, emitted from the vicinity of the first stars.
Since the ionization state and temperature of the gas around the first stars are determined by their luminosity, it is highly expected that the distribution of 21-cm line signal reflects the stellar mass. 
\citet{Chen2008} computationally showed that a distinctive 21-cm signature appears around the first star. 
In the very vicinity of the first star, there is almost no 21-cm signal because the gas is highly ionized. 
At the ionization front (I-front), the gas is partially ionized and heated. 
The gas far from the I-front is colder than the Cosmic Microwave Background (CMB) temperature, and the gas temperature almost equals to the spin temperature owing to the coupling process through the Lyman-$\alpha$ ($\rm Ly\alpha$) pumping \citep[Wouthuysen-Field effect; hereafter WF effect][]{Wouthuysen1952, Field1958}. 
Consequently, a 21-cm emitting region is surrounded by a 21-cm absorption region.
\citet{Yajima2014} also studied the 21-cm signature around the first stars by conducting radiative transfer simulations, and estimated the detectability of the signal with the Square Kilometre Array (SKA). 
They concluded that detecting the signal is difficult even with the SKA. 

However, these studies consider a static and uniform medium around the first stars and seem to miss some important points that affect the estimated 21-cm signature. 
For example, the dense gas distributed in a mini-halo, which they did not resolve, likely impacts the escape fraction of ionizing photons from the halo, thus the resultant size of the ionized region around the halo may change accordingly. 
In order to estimate the escape fraction precisely, radiation hydrodynamics (RHD) simulations are required.  
\citet{Kitayama2004} performed RHD simulations to study the evolution of ionized regions around the first stars. 
They found that a halo hosting the first star is completely photo-ionized if an I-front changes from the D-type I-front to the R-type I-front\footnote{A D-type I-front slowly propagates in a high-density region. On the other hand, an R-type I front travels through a rarefied medium with supersonic speed.} within the lifetime of the central star. 
Since the transition of the I-front occurs when the central region in a halo dynamically expands due to the thermal pressure enhanced by photo-heating, the type transition is sensitive to the amount of the gas in the halo, the potential of the halo, and the luminosity of the first star. 
Therefore the time evolution of the escape fraction strongly depends on the stellar and halo masses \citep{Kitayama2004}. 

In addition to the 21-cm signal in the vicinity of an individual halo, the sky-averaged 21-cm signal (the so-called global 21-cm signal) likely has some imprints reflecting the properties of the first stars. 
This was theoretically predicted \citep{Furlanetto2006, Pritchard2010, Mesinger2013, Yajima2015} and recently attracts more attention due to the claim of detection of the signal \citep{Bowman2018}.
However, previous studies based on simplified one-zone models 
did not take into account 
the stellar-mass dependence of the global signal; 
the stellar-mass-dependent escape fraction mentioned above brings about the mass-dependent-heating rate of the intergalactic medium (IGM). 
Besides, even with a given cosmic star formation rate density (SFRD), the cosmic stellar mass density would depend on the stellar mass function of the first stars because their lifetime has a dependence on their mass. 

To obtain reliable 21-cm signatures around the first stars, in this work, we conduct a series of spherically symmetric one-dimensional RHD simulations for individual first stars. 
In particular, we resolve the high-density region within a halo, by which we are able to appropriately consider the escaping process of ionizing photons from a halo and to evaluate time evolution of the radial profile of the 21-cm brightness temperature. 
Based on our simulation results, we explore the dependence of the 21-cm signal around the first stars on their stellar mass, halo mass, and their formation redshift. 
Furthermore, using our simulation results, we evaluate the detectability of the 21-cm signal around individual first stars and the dependence of the global 21-cm signal on properties of the first stars. 

This paper is organised as follows. 
In Section~\ref{sec:model}, we describe the methodology of our simulations.
Then in Section~\ref{sec:results}, we show simulated spatial distributions of the 21-cm brightness temperature around individual first stars and evaluate the detectability of these signals with the SKA. 
In Section~\ref{sec:global}, we compute the global 21-cm signal by using our simulation results.
Section~\ref{sec:discussion} is devoted to discuss uncertainties in our results.
Finally, we summarize our study in Section~\ref{sec:conclusion}. 

Throughout this paper, we work with a flat $\Lambda$CDM cosmology with the matter density $\Omega_{\mathrm{m}0} = 0.308$, the Hubble constant $h_0=0.678$, and the baryon density $\Omega_{\mathrm{b}0} = 0.0485$ (Planck Collaboration XIII \citeyear{PlanckXIII}).

\section{Mothod}\label{sec:model}
\subsection{Setup}
We consider a single first star embedded in a mini-halo, and each run starts just after the ignition of the first star. 
Each simulation run is characterized by three parameters: the halo mass ($M_{\rm halo}$), the stellar mass ($M_{\rm star}$), and the formation redshift of the first star ($z_{\rm f}$). 
The halo mass is defined as the total mass of three components, i.e, the gaseous, stellar, and dark matter components, within the virial radius $r_\mathrm{vir}$: 
\begin{equation}
	M_\mathrm{halo} = M_\mathrm{star} + M_\mathrm{gas} +M_\mathrm{DM},
\end{equation}
\begin{equation}
	M_\mathrm{DM} = \left( 1-\frac{\Omega_\mathrm{b}}{\Omega_\mathrm{m}} \right)
	M_\mathrm{halo},
\end{equation}
where $M_\mathrm{gas}$ and $M_\mathrm{DM}$ are respectively the gas mass and the dark matter mass. 
Once the halo mass is determined, the corresponding virial radius can be expressed as
\begin{equation}
	r_\mathrm{vir} = 
	50.8 \left( \frac{M_\mathrm{halo}}{10^5 \mathrm{M}_{\sun}} \right)^{1/3}
	\left( \frac{1+z_{\rm f}}{20} \right)^{-1}~\rm physical~pc.
\end{equation}

The initial gas density profile in a halo is assumed to obey a power-law distribution with the index of $-2.2$, which is indicated by previous studies \citep[e.g.,][]{Omukai1998, Susa2014, Hirano2015}, and is serially connected to the uniform IGM density at a given redshift, $\rho_\mathrm{IGM}(z)=8.6\times10^{-30}\Omega_\mathrm{b0}(1+z)^3$ g cm$^{-3}$. 
Hence the gas density distribution is given by 
\begin{equation}
	\rho_\mathrm{gas}(r) = \max\left\{\rho_{\rm IGM}(z_{\rm f}), ~\rho_\mathrm{gas,c} \left(\frac{r}{r_\mathrm{c}} \right)^{-2.2}~\right\}, 
\end{equation}
where $r$ is the distance from the centre of the halo, $\rho_\mathrm{gas,c}$ is the gas mass density of the innermost shell, and $r_\mathrm{c}$ is the core radius. 
We set $\rho_\mathrm{gas,c} = 1.67\times 10^{-18}~\rm g~cm^{-3}$ so that the central gas number density corresponds to $10^6~\rm {cm^{-3}}$, as in \citet{Kitayama2004}. 
The core radius, $r_\mathrm{c}$, is determined to satisfy a condition of $4\pi \int_0^{r_{\rm vir}} \rho_{\rm gas}(r) r^2 \mathrm{d}r = M_{\rm gas}$. 
As for the dark matter density profile in a halo,
we use the NFW profile \citep{Navarro1995,Navarro1996,Navarro1997} given by 
\begin{equation}
	\rho_\mathrm{DM}(r) =
	\frac{\rho_{0,\mathrm{DM}}}
	{(r/r_\mathrm{s}) (1+r/r_\mathrm{s})^2},
\end{equation}
where $r_\mathrm{s}$ is the scale radius defined as $r_\mathrm{s} = r_\mathrm{vir}/c_\mathrm{vir}$. 
We employ the concentration parameter, $c_\mathrm{vir}$, depending on redshift and halo mass shown by \citet{Bullock2001}. 
The amplitude of the profile is determined so that  the total dark matter mass within $r_{\rm vir}$ corresponds to $M_\mathrm{DM}$. 

The initial gas temperature is determined based on the following assumption;  
The gas adiabatically evolves after the decouple epoch $1+z_\mathrm{dec}\approx 137$~\citep{Peebles1993}, and the gas in high-density regions cool via molecular hydrogen cooling down to $500$~K. 
Letting $T_{\rm IGM}(z_{\rm f})$ be the IGM temperature at $z_{\rm f}$, the temperature distribution is given by
\begin{eqnarray}
	T_\mathrm{gas}(r) ~=  
	\min \left\{
	T_{\rm IGM}(z_{\rm f})
	\left( \frac{\rho_\mathrm{gas}(r)}{\rho_\mathrm{IGM}(z_{\rm f})}\right)^{2/3},
	500~\mathrm{K} \right\}, 
	\label{eq:tgas}
\end{eqnarray}
where $T_{\rm IGM}(z_{\rm f})=T_{\mathrm{CMB},z_\mathrm{dec}} \left( \frac{1+z_{\rm f}}{1+z_\mathrm{dec}}\right)^2$, with being $T_{\mathrm{CMB},z_\mathrm{dec}} = 2.725\times(1+z_\mathrm{dec})$~K at the decoupling epoch $z_\mathrm{dec}$. The index of $2/3$ is derived from the assumption that the gas is adiabatically heated up.

For the initial velocity, we only consider the Hubble velocity, $v_\mathrm{init}(r)$ $= r/H(z_\mathrm{f})$. 
In reality the gas in a halo should still accretes toward the centre of the halo when the central star is born. 
However, such an initial infall velocity hardly changes the resultant structure of the 21-cm brightness temperature because it reverses due to the thermal pressure of ionized gas soon after the simulation starts. 
 
\begin{table*}
\begin{center}
  \begin{tabular}{cccccccc}\hline
  Run Name & Dynamics & Halo gas & Redshift & Stellar Mass~[$\mathrm{M}_{\sun}$] & Halo Mass~[$\mathrm{M}_{\sun}$] & $t_{\rm life}$~[Myr] \\  \hline
  Run-Ref & no & no & $20$ & $100$ & - & 2.7  \\
  Run-z20S100H8e5 & yes & yes & $20$ & $100$ & $8\times10^5$  & 2.7 \\ 
  Run-z20S500H8e5 & yes & yes & $20$ & $500$ & $8\times10^5$ & 1.9 \\ 
  Run-z20S40H8e5 & yes & yes & $20$ & $40$ & $8\times10^5$ & 3.8 \\ 
  Run-z20S100H3e5 & yes & yes & $20$ & $100$ & $3\times10^5$ & 2.7 \\ 
  Run-z20S100H3e6 & yes & yes & $20$ & $100$ & $3\times10^6$ & 2.7 \\ 
  Run-z10S100H8e5 & yes & yes & $10$ & $100$ & $8\times10^5$  & 2.7 \\ 
  Run-z30S100H8e5 & yes & yes & $30$ & $100$ & $8\times10^5$ & 2.7 \\ 
  \hline
  \end{tabular}
  \end{center}
  \caption{Summary of representative runs performed in this paper. Note that each run is named after the formation redshift, the stellar mass, and the halo mass employed in the run. Dynamics means whether each run includes gas dynamics or not,
Halo gas indicates whether we consider halo gas whose density profile is explained in Section~\ref{sec:model},
redshift is the redshift at which the star is formed and begins to emit photons. \label{tb:param}}
\end{table*}

We conduct each simulation during the main-sequence lifetime of the first star, $t_\mathrm{life}$. 
The values of the effective temperature, $T_\mathrm{eff}$,  the total number of ionizing photon emitted per second, $\dot{n}_\mathrm{ion}$, and $t_\mathrm{life}$, depending on the stellar mass $M_{\rm star}$, are taken from \citet{Schaerer2002}. 
For simplicity, the time evolution of the spectrum is not considered. 

Since it is important to discuss the impact of resolving dense gas in a mini-halo, we also perform a reference run employing a static and uniform medium, which is assumed in previous studies \citep{Chen2008, Yajima2014}.
In this run, we assume $50$ per cent of ionizing photons can escape from the halo. 
Table~\ref{tb:param} summarizes representative runs that we perform in this study.

\subsection{Simulation code} \label{simulationcode} 
Based on the initial setup, the evolution of the gas in a halo is calculated with a modified version of an RHD simulation code used in previous studies \citep{Kitayama2000, Kitayama2001, Kitayama2004, HUK2009}, which adopts the Lagrangian finite-difference scheme in a spherically symmetric geometry. 
The code enables us to solve hydrodynamics, non-equilibrium chemistry regarding primordial gas, and the radiative transfer of ionizing photons self-consistently. 

The basic equations for a Lagrangian gas shell are 
\begin{equation}
	\frac{\mathrm{d}m}{\mathrm{d}r} = 4\pi r^2 \rho,
\end{equation}
\begin{equation}
	\frac{\mathrm{d}^2 r}{\mathrm{d}t^2} = -4\pi r^2 \frac{\mathrm{d}p}{\mathrm{d}m} - 
	\frac{GM_\mathrm{tot}(r)}{r^2} + \frac{\mathrm{d}(rH)}{\mathrm{d}t},
	\label{eq:motion}
\end{equation}
\begin{equation}
	\frac{\mathrm{d}u}{\mathrm{d}t} = \frac{p}{\rho^2} \frac{\mathrm{d}\rho}{\mathrm{d}t} +
	\frac{\mathcal{H}-\mathcal{L}}{\rho},
	\label{eq:energy}
\end{equation}
where $r$, $m$, $\rho$, $p$, $u$, $M_\mathrm{tot}(r)$, $H$, $\mathcal{H}$, and $\mathcal{L}$ are the distance from the central star, mass, mass density, pressure, specific internal energy, the total mass inside $r$ of the shell, the Hubble parameter, the heating rate, and the cooling rate, respectively. Also, $m_{\rm p}$ and $k_{\rm B}$ respectively indicate the proton mass and the Boltzmann constant. 
The equation of state, $p = (2/3)\rho u = \rho k_{\rm B}T_{\rm gas}/(\mu m_{\rm p})$, is used to close the equations above, letting $\mu$ be the mean molecular weight. 
We treat the dark matter component as a static medium, by which we consider a dynamically relaxed (virialized) halo. 
As for the equation of motion, equation~(\ref{eq:motion}), we neglect the radiation force caused by ionizing photons but consider the Hubble expansion, because of the following reasons: 
(i) the impact of the radiation force on the gas in a halo is remarkable only at a very early phase after the central first star starts to shine \citep{Kitayama2004}, and (ii) the 21-cm signal around the first star is often extended up to $1$ comoving Mpc where the Hubble expansion is not negligible \citep[e.g.,][]{Chen2008,Yajima2014}. 

To solve the energy equation, equation~(\ref{eq:energy}), the heating rate, $\cal{H}$, and the cooling rate, $\cal{L}$, are obtained by solving the radiative transfer of ionizing photons and the chemical reactions regarding primordial gas.
The chemical evolution follows 
\begin{equation}
	\frac{\mathrm{d} n_i}{\mathrm{d}t} = C_{i}(T_{\rm gas}, n_j) - D_i (T_{\rm gas}, n_j) n_i, 
\end{equation}
where $n_i$ is the number density of the $i$-th species, $C_{i}$ and $D_{i}$ are the creation and destruction rates of the $i$-th species. 
In addition to chemical species of $\rm e$, H\textsc{i}, H\textsc{ii}, H$^-$, H$_2$, H$_2^+$, which are originally considered in \citet{Kitayama2004}, we further consider He\textsc{i}, He\textsc{ii}, and He\textsc{iii} to calculate the distributions of H\textsc{i} and gas temperature accurately.  
The chemical reactions and cooling rates of He are taken from \citet{Fukugita1994}. 
The thermal and chemical evolution are implicitly solved at the same time, assuming the initial abundance shown by \citet{Galli1998}. 

When we solve the thermal and chemical evolution, the photo-heating and photo-ionization rates of $i$-th species ($i$=H\textsc{i}, He\textsc{i}, and He\textsc{ii}) should be determined consistently, by solving the radiative transfer. 
Taking the weight of the absorption probability by the $i$-th~species into account, we estimate the photo-ionization rates at each time step as \citep{Susa2006a, Yoshiura2017}, 
\begin{align}
	k_{i,\gamma}(r) = \frac{1}{V_\mathrm{shell}n} \int^\infty_{\nu_{i}}\mathrm{d}\nu
	\frac{n_i \sigma_i(\nu)}{\sum_j n_j \sigma_j(\nu)} \frac{L(\nu)}{h\nu}
	\mathrm{e}^{-\tau_\nu(r)} 
	(1-\mathrm{e}^{\Delta\tau_\nu(r)}) ,
\end{align}
where $V_\mathrm{shell}$ is the volume of the shell, $\sigma_i(\nu)$ is the cross section of the $i$-th~species, $L(\nu)$ is the luminosity of the central source, $\nu_i$ is the ionization threshold frequency of the $i$-th species, $\tau_\nu(r)$ is the optical depth between the central star and the shell,
\begin{equation}
	\tau_\nu(r) = \sum_i \sigma_{i}(\nu) \int^{r}_0 n_i \mathrm{d}r, 
\end{equation}
$\Delta\tau_\nu$ is the optical depth in the shell, and $h$ is the Planck constant. 
In the radiative transfer calculation, we employ the on-the-spot approximation because the transfer of diffuse photons hardly affects the dynamics and the distribution of neural hydrogen \citep{Hasegawa2010}. 

Similar to the photo-ionization rates, the total photo-heating rate is given by 
\begin{align}
	\mathcal{H}(r) = \frac{1}{V_\mathrm{shell}}\sum_i \int^\infty_{\nu_{i}} \mathrm{d}\nu & 
	\frac{n_i \sigma_i(\nu)}{\sum_j n_j \sigma_j(\nu)} \frac{L(\nu)}{h\nu}
	\mathrm{e}^{-\tau_\nu(r)}\nonumber \\
	&\times (1-\mathrm{e}^{\Delta\tau_\nu(r)}) (h\nu - h\nu_i) (1-\eta_\alpha),
\end{align}
where, $\eta_\alpha$ is the fraction of the ejected electrons' energy which is not directly converted to the thermal energy but is used for exciting neutral hydrogen (see Section~\ref{calc21cm}). 
We use the formula $\eta_\alpha(x_{\mathrm{H}\textsc{ii}})$ = 0.4766(1$-$ $x_{\mathrm{H}\textsc{ii}}^{0.2735}$)$^{1.5221}$, where $x_{\mathrm{H}\textsc{ii}}$ is the ionization fraction of hydrogen \citep{Shull1985}.

Since the relevant physical scales widely range from $\sim0.1\rm ~pc$~(the core scale of a halo) to $\sim100~\rm kpc$ (the size of an ionized bubble), a uniform shell mass is inappropriate to solve the equation of motion. 
Therefore, we employ the shell mass that logarithmically increases toward outside,  
\begin{equation}
	\log m_n = 
	\log m_\mathrm{min}	 + (n-1) \frac{\log m_\mathrm{max}-\log m_\mathrm{min}}{N_\mathrm{bin}},
\end{equation}
where, $m_{n}$ is the mass of the $n$-th shell ($n = 1, 2, \cdot \cdot \cdot, N_\mathrm{bin} $), $N_\mathrm{bin}$ is the total number of the shells for which we adopt $N_\mathrm{bin} = 500$. 
With $m_\mathrm{min}$ being the central mass of a halo ($4\pi/3 r_\mathrm{c}^3 \rho_\mathrm{gas,c}$) and $m_\mathrm{max} = 10^{10} \mathrm{M}_{\sun}$,  
we can resolve inner $\sim0.1~\rm physical~pc$, and can trace the evolution of an ionization front up to $\sim 100~\rm physical~kpc$. 

\subsection{Computing 21-cm signal}\label{calc21cm}
Based on results from the RHD simulations, we compute the 21-cm signal around the first star. 
We start with a brief introduction of the cosmological 21-cm signal and then introduce our computing method.

The cosmological H\textsc{i} 21-cm signal can be measured as the difference from the CMB radiation. 
Assuming the peculiar velocity of gas is negligible compared to the Hubble velocity, the differential brightness temperature at a given redshift, $z$, is estimated by \citep[e.g.,][]{Furlanetto2006Oh}
\begin{align}
	\delta T_\mathrm{b} =& \frac{T_{\rm S} - T_{\rm CMB}(z)}{1+z}(1-{\rm e^{-\tau_{21}}}) \nonumber \\
	\approx & 38.7
	\frac{n_\mathrm{H\textsc{i}}}{\bar{n}_\mathrm{H}}
	\left( \frac{1+z}{20} \right)^{1/2}
	\frac{T_\mathrm{S}-T_\mathrm{CMB}(z)}{T_\mathrm{S}} ~\mathrm{mK},
	\label{eq:dtb}
\end{align}
where $T_{\rm S}$, $T_\mathrm{CMB}(z)$, $\tau_{\rm 21}$, $n_{{\mathrm{H}\textsc{i}}}$, and $\bar{n}_\mathrm{H}$ are the spin temperature, the CMB temperature at a given redshift $z$, the optical depth for the H\textsc{i} 21-cm line, the number density of neutral hydrogen, and the average number density of hydrogen nuclei in the Universe, respectively. 
The spin temperature is determined by the excitation and de-excitation by the CMB photons, collisions with atoms and electrons, and the pumping by $\rm Ly\alpha$ photons (the WF effect), thus given by 
\begin{equation}
	T_\mathrm{S}^{-1} = \frac{
	T_\mathrm{CMB}^{-1}+x_\mathrm{c}T_\mathrm{gas}^{-1}+ x_\alpha T_\alpha^{-1}}{
	1 + x_\mathrm{c} + x_\alpha},
	\label{eq:tspin}
\end{equation}
where $T_\alpha$ is the colour temperature of $\rm Ly\alpha$ photons.
As far as we consider optically thick gas, the assumption of $T_\alpha = T_\mathrm{gas}$ is generally valid because a number of $\rm Ly\alpha$ photons enable the colour temperature $T_\alpha$ to coincide with the gas temperature $T_\mathrm{gas}$.
The collisional coupling coefficient, $x_\mathrm{c}$, and the $\rm Ly\alpha$ coupling coefficient, $x_\mathrm{\alpha}$, are respectively given by 
\begin{equation}
	x_\mathrm{c} = 
	\frac{C_{10}T_*}{A_{10}T_\mathrm{CMB}},
	\ \ \ \ \ \ \ 
	x_\mathrm{\alpha} =
	\frac{P_{10}T_*}{A_{10}T_\mathrm{CMB}},
\end{equation}
where $C_{10}$, $P_{10}$, $A_{10}$, and $T_{\ast}$ are the de-excitation rate due to collisions, the number of $\rm Ly\alpha$ photon scattering per atom per unit time, the spontaneous emission coefficient of the hyperfine structure of neutral hydrogen, and the temperature corresponding to the transition energy. 
For the H\textsc{i} 21-cm line, $T_* = 0.068$~K and $A_{10}= 2.85 \times 10^{-15}$~s$^{-1}$. 

In order to determine $C_{10}$, we take into account collisions of neutral hydrogen with neutral hydrogen, protons and electrons: 
\begin{equation}
	C_{10} = n_{\mathrm{H}\textsc{i}}\kappa_{\mathrm{H}\textsc{i}}(T_{\rm gas}) + n_{\rm e}\kappa_{\rm e}(T_{\rm gas}) + n_{\rm p}\kappa_{\rm p}(T_{\rm gas}).
\end{equation}
As for these collisional coupling coefficients, $\kappa_{\mathrm{H}\textsc{i}}$, $\kappa_{\rm e}$, and $\kappa_{\rm p}$, 
we use the fitting formulae given by \citet{Kuhlen2006}, \citet{Liszt2001}, and \citet{Smith1966}. 
The quantities required for calculating the coefficients, i.e. $n_{\mathrm{H}\textsc{i}}$, $n_{\rm e}$, $n_{\rm p}$, and $T_{\rm gas}$, can be obtained from the RHD simulations. 

We basically follow the method developed by \citet{Chen2008} to calculate the coupling via the WF effect. 
The $\rm Ly\alpha$ mean intensity 
at a position $r$ is required to obtain $P_{10}$, for which we consider two processes. 
One is the redshifted ultraviolet (UV) continuum photons emitted from the central star with frequencies between the $\rm Ly\alpha$ frequency and the Lyman limit frequency: 
The resonance occurs when the redshifted photons coincide with particular energies of the Lyman series. 
In this case, the subsequent cascade process leads to the production of new $\rm Ly\alpha$ photons. 
The number intensity of these ``recycled" $\rm Ly\alpha$ photons can be written by
\begin{equation}
	J_\mathrm{c} =
	\sum_{n=2}^{n_\mathrm{max}} \Theta (\nu_{n+1} - \nu'_n)
	f_\mathrm{recycle}(n) \frac{N(\nu'_n)}{(4\pi)^2 r^2},
\end{equation}
where $\Theta$ is the Heaviside function to take into account horizon scales of Lyman series photons \citep{Ahn2015}, $\nu'_n$ is the frequency of emitted photons at the rest frame of the central star, which can redshift to the frequency corresponding the $n$-th excitation level, $\nu_n$, when they arrive at $r$, 
$N(\nu)$ is the number luminosity per frequency of the central star, and  $f_\mathrm{recycle}(n)$ is the fraction of Lyman series photons which turn out to be $\rm Ly\alpha$ photons via cascades \citep{Furlanetto2006Oh}. 
The  typical optical depth of each shell in our simulations is much larger than unity. For that reason, we assume that
emitted photons with frequencies of $\nu_n$-$\nu_{n+1}$ are immediately absorbed when their frequencies becomes $\nu_n$. 
Therefore, each Lyman series photon has its horizon scale, beyond which any absorptions do not happen. 
The summation of $n$ should be stopped at some large $n_\mathrm{max}$ for which we set $n_\mathrm{max}=30$ \citep{Furlanetto2006Oh}. 

The other process is excitations of neutral hydrogen by secondary electrons ejected via photo-ionization.  
Since we consider photons with hard spectra from the central stars, the ejected electrons have kinetic energies high enough to excite neutral hydrogen. 
The $\rm Ly\alpha$ mean intensity 
produced by subsequent de-excitations is expressed as 
\begin{equation}
	J_\mathrm{i}=
	\frac{c\eta_\alpha\mathcal{H}}{4\pi H(z) h \nu_\alpha^2}. 
	\label{eq:Ji}
\end{equation}

By using $J_{\rm c}$ and $J_{\rm i}$, $P_{10}$ is described as 
\begin{equation}
    \label{P10}
	P_{10} = \frac{4}{27} H(z) \tau_\mathrm{GP}(z)
	\frac{J_\mathrm{c}S_\mathrm{c} + 
	J_\mathrm{i}S_\mathrm{i}}{\tilde{J}_0}. 
\end{equation}
Here $\tilde{J}_0$ is defined as $c n_\mathrm{H}/(4\pi\nu_\alpha)$. The Gunn-Peterson optical depth, $\tau_\mathrm{GP}$, is given by
\begin{equation}
	\tau_\mathrm{GP} = \frac{\chi_\alpha \bar{n}_{\mathrm{H}\textsc{i}}(z) c}{H(z) \nu_\alpha}.
\end{equation}
Here, 
$\chi_\alpha = (\pi e^2/m_e c)f_\alpha$ 
is the $\rm Ly\alpha$ absorption cross section at the line centre , $f_\alpha=0.4162$ is the oscillator strength of the $\rm Ly\alpha$ transition, $\bar{n}_{\mathrm{H}\textsc{i}}$ is the average number density of the neutral hydrogen. 
In equation~(\ref{P10}), $S_\mathrm{c}$ and $S_\mathrm{i}$ are the suppression factors originated from the fact that the shape of the radiation spectrum changes during multiple scattering. 
We assume that $S_\mathrm{c} = S_\mathrm{i}$, and use the fitting formula given by \citet{Furlanetto2006Pritchard}:
\begin{equation}
	S_\mathrm{c} = S_\mathrm{i} \approx
	1-\frac{4\pi}{3\sqrt{3}\Gamma(2/3)}\alpha + \frac{8\pi}{3\sqrt{3}\Gamma(1/3)}\alpha^2
	- \frac{4}{3}\alpha^3,
\end{equation}
\begin{equation}
	\alpha = 0.717T_\mathrm{gas}^{-2/3} \left(\frac{\tau_\mathrm{GP}}{10^6}\right)^{1/3},
\end{equation}
in the wing approximation of the Voigt profile. Here, $\Gamma(x)$ is the Gamma function.

\section{21-cm signature around the first star}\label{sec:results}
\subsection{Importance of resolving gas in a halo}\label{sec:halogas}
\begin{figure}
	\begin{center}
	\includegraphics[bb = 0 0 720 1008, width=80mm]{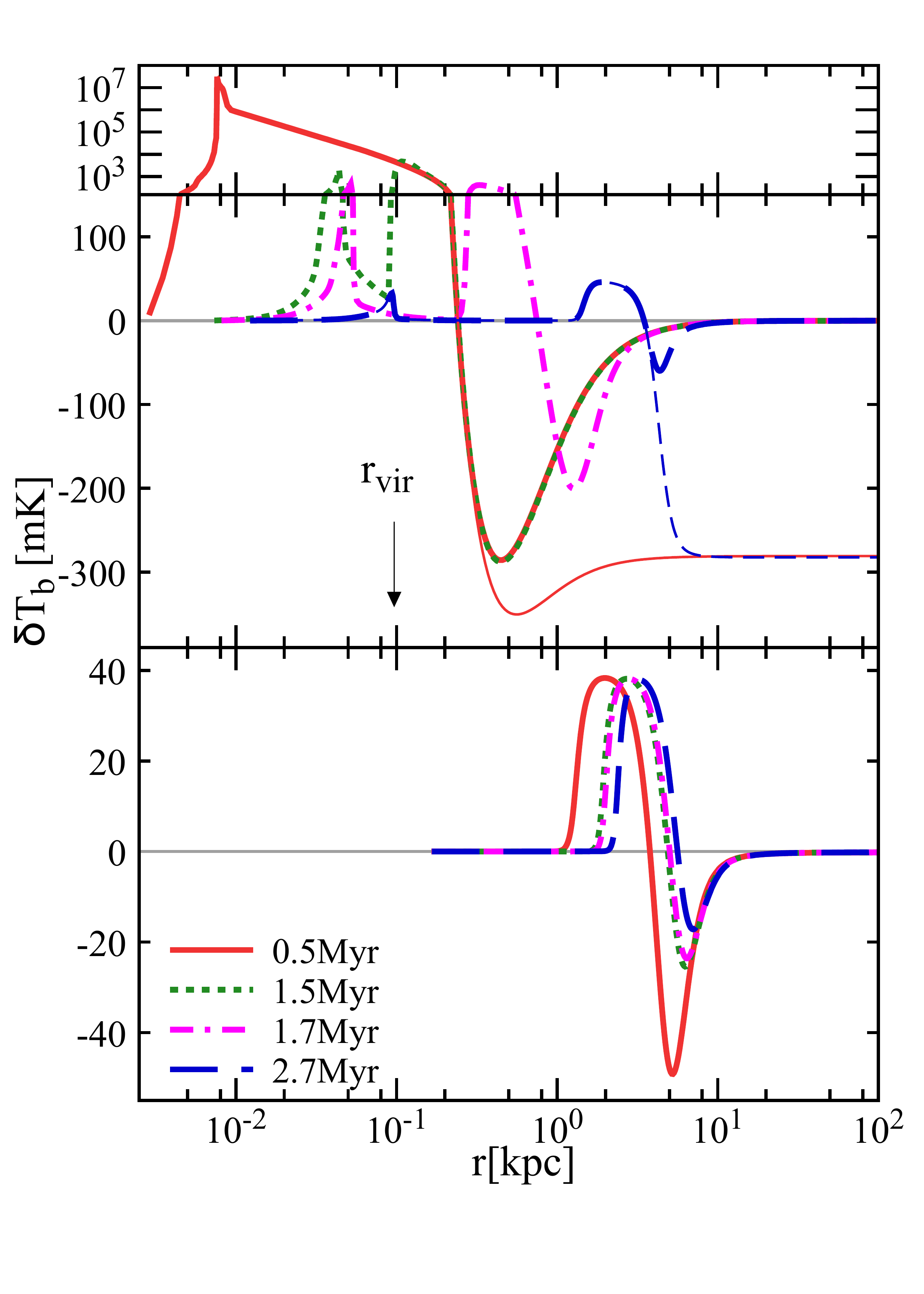}
 	\end{center}
	\caption{Radial profile of the differential brightness temperature as a function of the distance 
	from the central star.  
	The upper and lower panels respectively show the profiles in Run-z20S100H5e8 and Run-Ref. 
	In these runs, the parameters are $M_\mathrm{star} = 100 \mathrm{M}_{\sun}$, $z_{\rm f} = 20$, 
	and $M_\mathrm{halo} = 8\times10^5 \mathrm{M}_{\sun}$. 
	The red solid, green dotted, magenta dot-dashed, and blue dashed curves represent the profiles at $t_\mathrm{age} = $ $0.5
	 \mathrm{Myr}$, $1.5 \mathrm{Myr}$, $1.7 \mathrm{Myr}$, and $2.7 \mathrm{Myr}$, respectively. 
	The thin curves show $\delta T_{\rm b}(r)$ which is based on an assumption that the spin temperature is completely coupled with 
	the gas temperature. 
	The vertical arrow with $r_{\rm vir}$ indicates the virial radius. 
	}
	\label{fig1}
\end{figure}
\begin{figure*}
	\begin{center}
           	\includegraphics[bb = 0 0 1440 1728, width=310pt]{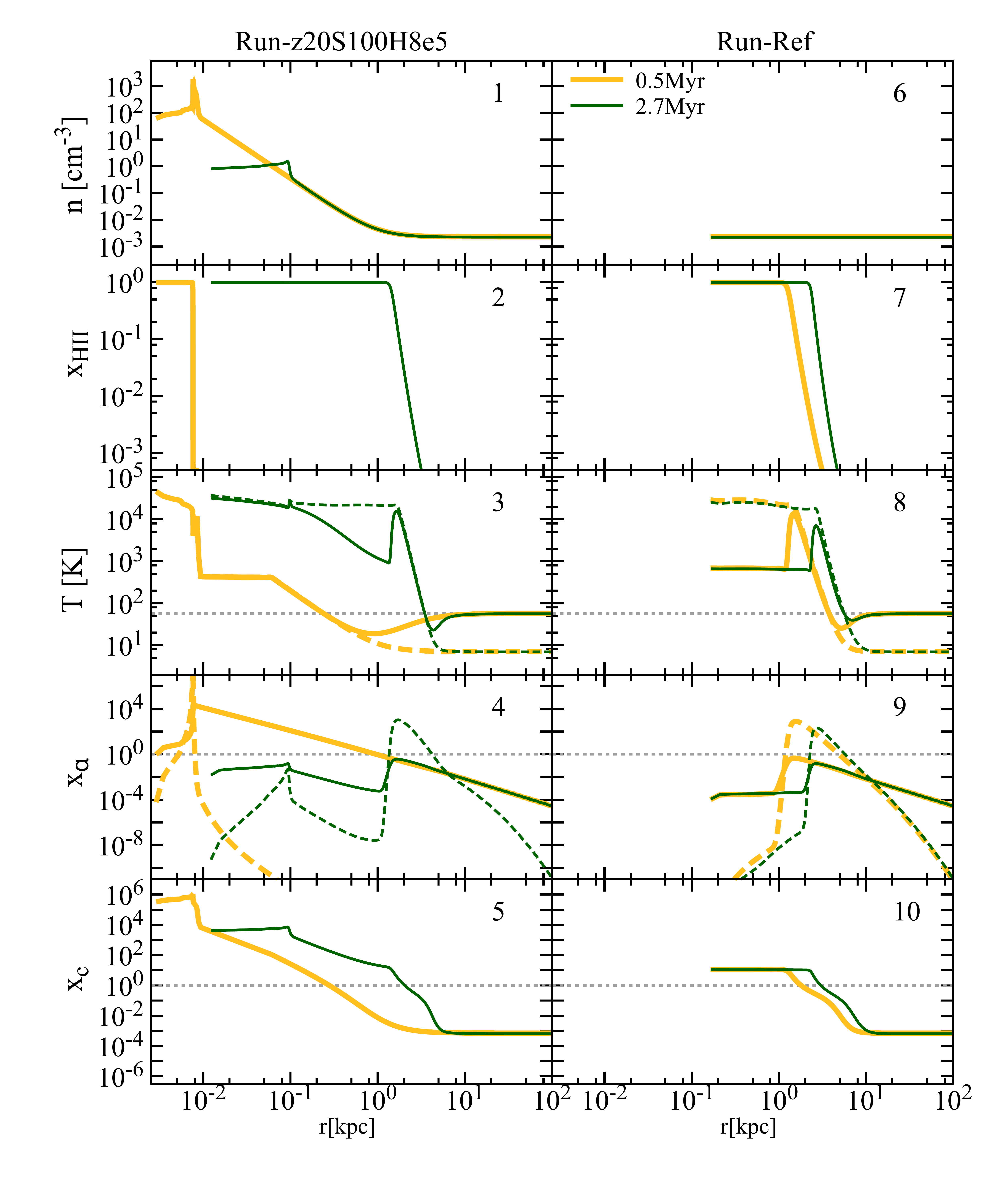}
	\end{center}
	\caption{
	Radial profiles of physical quantities associated with the differential brightness temperature 
	in Run-z20S100H8e5 (left column) and Run-Ref (right column) at $t_\mathrm{age} = 0.5 \mathrm{Myr}$ 
	(yellow thick curves) and $t_\mathrm{age} = 2.7 \mathrm{Myr}$ (green thin curves). 
	From top to bottom, each panel shows the radial profiles of the number density, ionized fraction of 
	hydrogen, the spin and gas temperatures, the  $\rm Ly\alpha$ coupling coefficients originated in
	the continuum photons (solid) and the secondary excitation (dashed), and the collisional coupling coefficient. 
	In Panel 3 and 8, the CMB temperature is shown by the horizontal dotted line, and the gas and spin 
	temperatures are respectively represented by the solid and dashed curves. 
	In Panel 4, 5, 9 and 10, the horizontal dashed line indicates unity above which the coupling processes 
	effectively work to couple the spin temperature to the gas temperature. 
	}
	\label{fig2}
\end{figure*}

\begin{figure}
	\begin{center}
	\includegraphics[bb = 0 0 504 504, width=60mm]{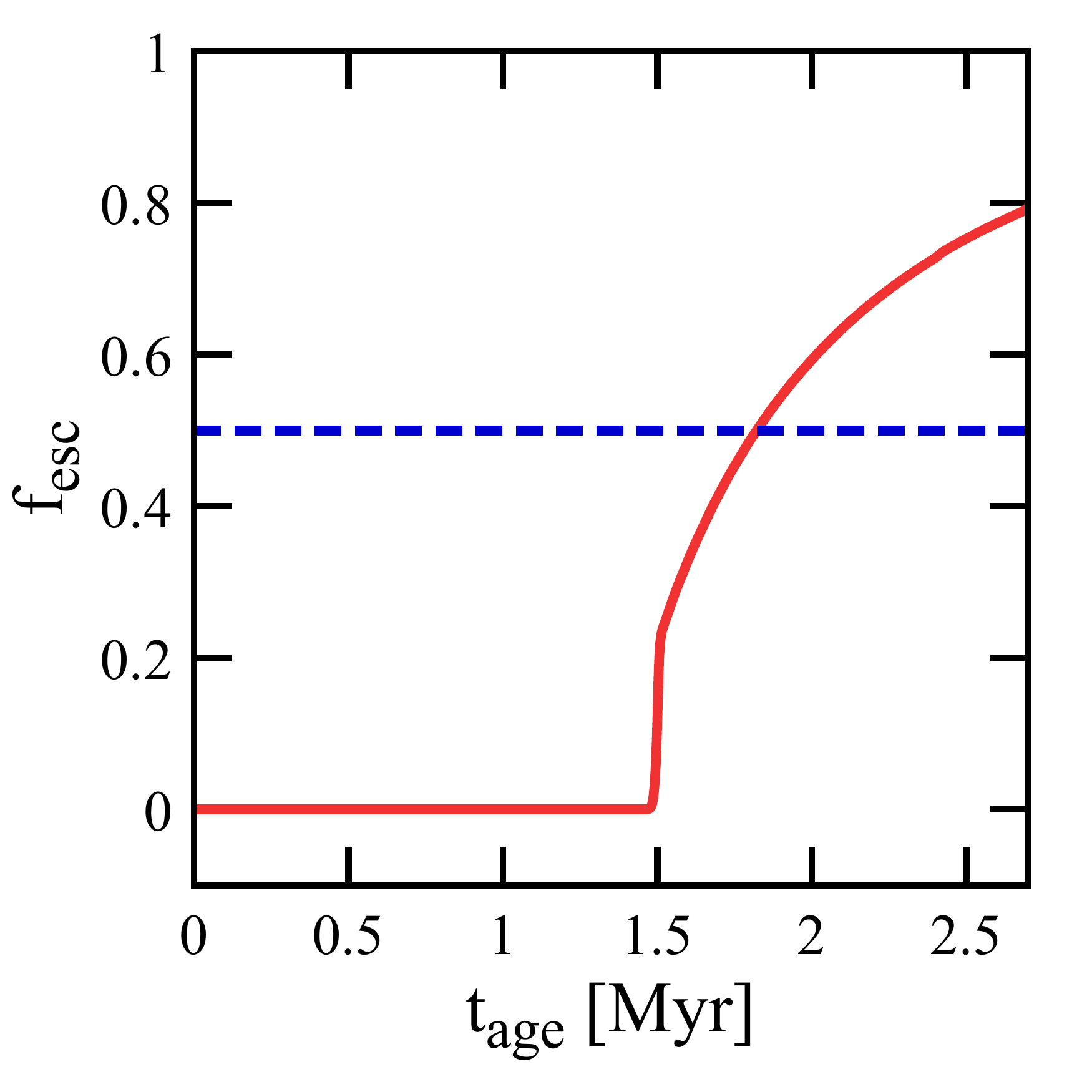}
	\end{center}
	\caption{
	Time evolution of the escape fraction in Run-z20S100H8e5. 
	The constant escape fraction of $f_\mathrm{esc} = 0.5$, which is assumed in Run-Ref, is also 
	shown by the horizontal dashed line. 
	}\label{fig3}
\end{figure}
As described in Section~\ref{intro}, considering gas dynamics in each halo is indispensable to precisely calculate the time-evolving escape fraction.
Thus, our simulations take into account gas dynamics and resolve gas distribution in a mini-halo unlike previous studies \citep{Chen2008, Yajima2014}. 
In this section, we investigate how the effects of resolving dense gas in a halo and gas dynamics impact the emerging radial distribution of $\delta T_{\rm b}$. 
To do this, we compare Run-Ref, in which static and uniform gas density is assumed, and Run-z20S100H5e8, in which the RHD effect is considered (see Table~\ref{tb:param}). 
Fig.~\ref{fig1} shows the radial profiles of the 21-cm brightness temperature at $t_{\rm age}=0.5$, $1.5$, $1.7$, and $2.7$~Myr, where the upper and lower panels correspond to the profiles obtained from Run-z20S100H5e8 and Run-Ref. 
Fig.~\ref{fig2} shows the time evolution of physical quantities relevant to the brightness temperature in these runs. 

In Run-Ref, a well-known characteristic feature, i.e. an emitting region surrounded by an absorption region, monotonically moves away from the central star as an I-front propagates. 
This characteristic feature is known to be caused by combination of partially ionized warm gas in the I-front and abundant $\rm Ly{\alpha}$ photons, forcing the spin temperature to be coupled with the gas temperature, in front of the I-front
(see the right panels in Fig.~\ref{fig2}).

Compared to Run-Ref, there are two remarkable features in Run-z20S100H5e8. 
One is the strong emission region at $r\sim10^{-2}~\rm kpc$, which disappears in a short time-scale within $0.5~\rm Myr$.
The other is the deep absorption feature outside the virial radius, which is almost steady until $t_{\rm age}=1.5~\rm Myr$ and then decays with time. 
Fig.~\ref{fig2} reveals that the spike in the emitting region ($\delta T_{\rm b}>0$) is caused by a shock preceding the I-front (see Panel~1 of Fig.~\ref{fig2}). 
Although the brightness temperature at the spike is high at the very early phase, the spike rapidly disappears because the shock propagates outward quickly. 
On the other hand, the deep absorption feature comes from the gas residing in the outer rim of the halo where the gas is slightly overdense and colder than the CMB temperature. 
As shown by \citet{Kitayama2004}, the ionized region is well confined until the I-front is converted from D-type to R-type (Panel~2 of Fig.~\ref{fig2}). 
Therefore photo-heating does not work at the outer rim of the halo, and 
the gas is kept as the cold state (Panel~3 of Fig.~\ref{fig2}). 
In contrast to the ionizing photons, UV continuum photons, which enable
the spin temperature to be coupled with the gas temperature through the WF effect, can reach the outer edge of the halo (Panel~4 of Fig.~\ref{fig2}). 
Furthermore, the radial profile of $\delta T_{\rm b} \propto n_{\mathrm{H}\textsc{i}}\left(1-\frac{T_{\rm CMB}}{T_{\rm S}}\right)$ has the local minimum (the thin solid curve in the upper panel of Fig.~\ref{fig1}) in the case that the spin temperature is completely coupled with the gas temperature ($T_{\rm S}=T_{\rm gas}$). 
The combination of the local minimum of $n_{\mathrm{H}\textsc{i}}\left(1-\frac{T_{\rm CMB}}{T_{\rm S}}\right)$ and the dilution of the $\rm Ly{\alpha}$ flux results in the remarkable absorption feature at the outer rim of the halo. 

The rapid decay of the absorption feature starts when
the ionizing photons escape from the halo. 
The escape fraction of ionizing photons can be calculated by 
\begin{equation}
	f_\mathrm{esc} =
	\frac{\int_{\nu_{\rm H\textsc{i},L}}^\infty 
	\frac{L(\nu)}{h\nu} \mathrm{e}^{-\tau_\nu (r_{\rm vir})}\mathrm{d}\nu}{\int_{\nu_{\rm H\textsc{i},L}}^\infty \frac{L(\nu)}{h\nu}\mathrm{d}\nu},
\end{equation}
where $\nu_{\mathrm{H}\textsc{i},\mathrm{L}}$ is the Lyman limit frequency. 
Fig.~\ref{fig3} shows the time evolution of the escape fraction of ionizing photons. 
The escape fraction sharply rises at $t_\mathrm{age} \sim 1.5$~Myr, when the absorption feature starts to decay. 
Hereafter, we call this characteristic time as $t_{\rm decay}$.
As shown by Fig.~\ref{fig3}, the escape fraction in Run-z20S100H5e8 after $t_{\rm decay}$ is higher than that assumed in Run-Ref, i.e., $f_\mathrm{esc} = 0.5$. 
This is a reason why the I-front in Run-z20S100H5e8 almost catches up that in Run-Ref at $t_\mathrm{age} > t_{\rm decay}$ (e.g., $t_\mathrm{age} \sim 2.7$~Myr in Panel 2 and 7 of Fig.~\ref{fig2}). 
We note that the expansion of the ionized region due to hydrodynamics does not appear significantly because the recombination time is much shorter than the lifetime of the central star. 
  
  
It is worth to mention that the dominant process coupling the spin temperature with the gas temperature differs between Run-Ref and Run-z20S100H5e8.
Since the spin temperature deviates from the CMB temperature if the coupling coefficient (either $x_{\rm c}$ or $x_{\alpha}$) is greater than unity, the size of a signal region is roughly determined by the position where a coupling coefficient becomes unity. 
As shown by Panel~5 of Fig.~\ref{fig2}, the collisional coupling process is important only in the high-density region, and is usually less efficient than $\rm Ly\alpha$ coupling processes. 
Thus, we hereafter focus solely on $x_{\alpha}$.  
In Run-Ref, the size of the signal region is determined by the secondary excitation process rather than redshifted UV continuum at any evolutionary phases (see Panel 9 of Fig.~\ref{fig2}), because the absorption region moves with the I-front where electrons are ejected by photo-ionization \citep{Chen2008}. 
On the other hand, in Run-z20S100H5e8, the process determining the size of the signal region differs according to dynamical evolution. 
Before $t_{\rm decay}$, the size is determined by the redshifted UV continuum photons emitted from the central star, as shown in Panel 4 of Fig.~\ref{fig2}. 
The contribution from $\rm Ly\alpha$ flux induced by the secondary electrons is important only inside the halo, because the ionized region is confined in the halo.  
After $t_{\rm decay}$, $T_{\rm S}$ and $T_{\rm gas}$ become coupled via the  $\rm Ly\alpha$ photons associated with the secondary excitation as in Run-Ref. 

The deep absorption region may be an attractive target for observations in the future, because the existence of such a deep absorption implies a mini-halo hosting the first star. 
Even if the profile is not resolved, the spatially smoothed signal within the resolution of a telescope is more or less influenced by the high amplitude as we discuss in Section~\ref{sec:3.5} and Section~\ref{sec:3.6}. 
We also show that the decay time $t_\mathrm{decay}$ plays an important role on the detectability of the individual 21-cm signal, and is strongly dependent on the stellar mass, the halo mass, and the formation redshift in the following sections Section~\ref{sec:mstar}, Section~\ref{sec:mhalo}, and Section~\ref{sec:redshift}. 

\subsection{Stellar mass dependence of $\delta T_{\rm b}(r)$}\label{sec:mstar}
Since unveiling the mass of the first stars is very important issue, we now investigate how the 21-cm signal around the first star depends on the stellar mass. 

\begin{figure}
	\begin{center}
    \includegraphics[bb = 0 0 1296 1296, width=80mm]{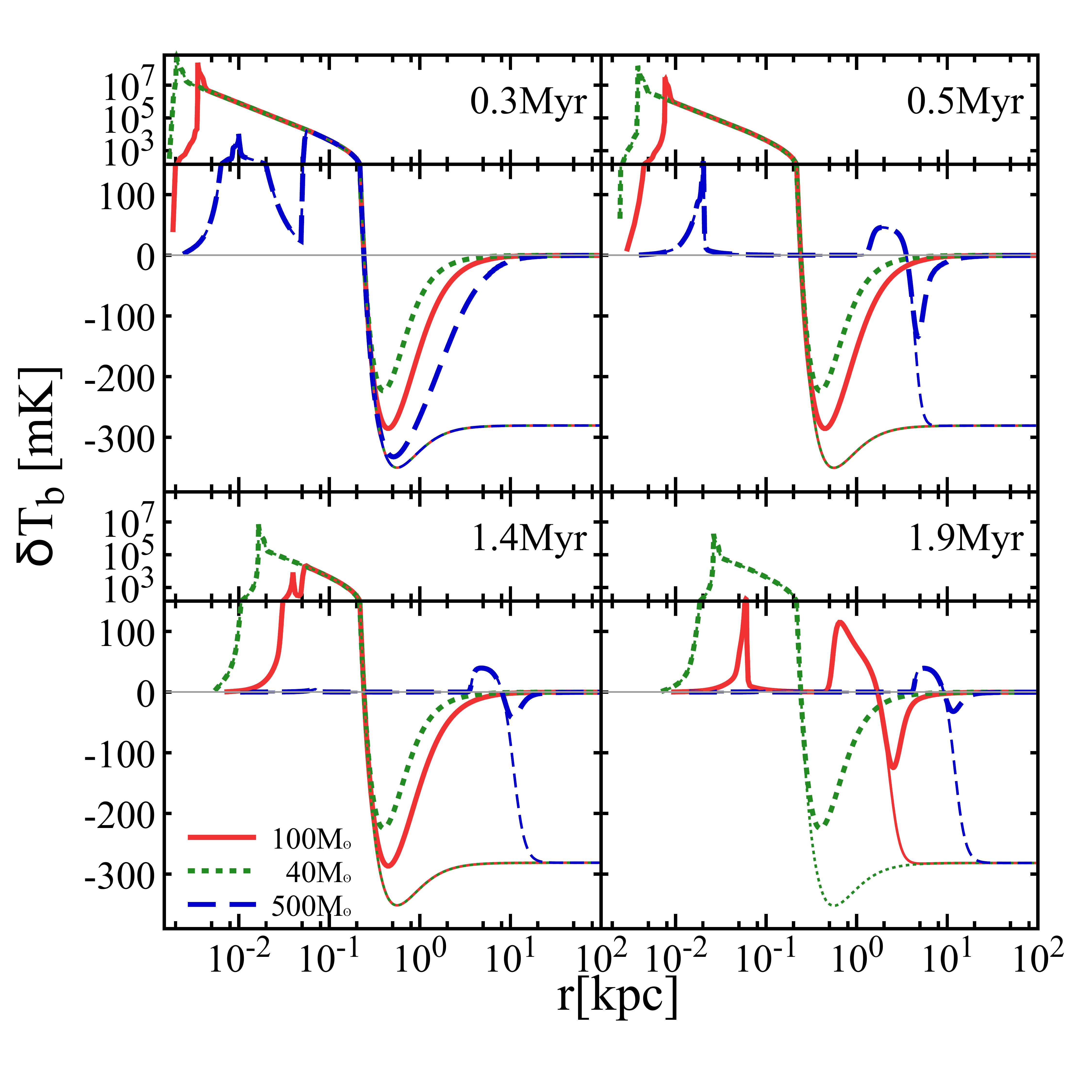}
	\end{center}
	\caption{
	Radial profiles of the differential brightness temperature at $
	t_\mathrm{age} = 0.3~\mathrm{Myr}$~(upper left), 
	$0.5~\mathrm{Myr}$~(upper right), $1.4~\mathrm{Myr}$~(lower left), 
	and $1.9~\rm Myr$~(lower right). 
	The profiles obtained from Run-z20S500H8e5, Run-z20S100H8e5, and Run-z20S40H8e5 are shown by 
	the blue dashed, red solid, and green dotted curves, respectively. 
	The thin curves show $\delta T_{\rm b}(r)$ assuming the spin temperature tightly coupled with
	the gas temperature. 
}\label{fig4}
\end{figure}
\begin{figure}
	\begin{center}
	\includegraphics[bb = 0 0 504 504, width=60mm]{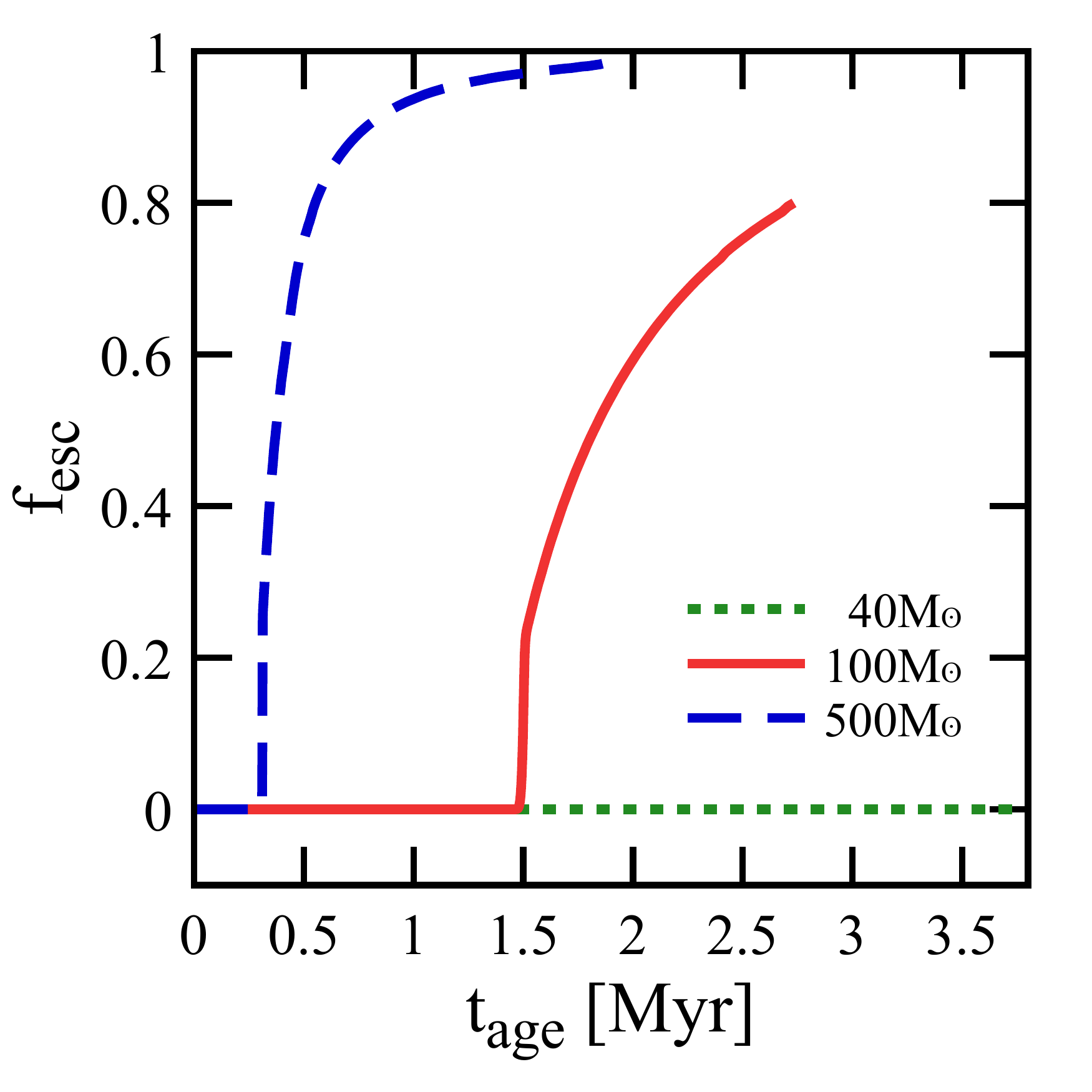}
	\end{center}
	\caption{
	Time evolution of the escape fraction in Run-z20S500H8e5~(blue dashed curve), 
	Run-z20S100H8e5~(red solid curve), and Run-z20S40H8e5~(green dotted curve). 	
	}\label{fig5}
\end{figure}

Fig.~\ref{fig4} shows the time evolution of the radial profiles of the brightness temperature in Run-z20S500H8e5, Run-z20S100H8e5, and Run-z20S40H8e5, in which stellar masses are respectively $500~\rm M_{\sun}$, $100~\rm M_{\sun}$, and $40~\rm M_{\sun}$. 
In the runs, other parameters are fixed as $M_\mathrm{halo} = 8\times 10^5 \mathrm{M}_{\sun}$ and $z_{\rm f} = 20$. 
At a very early phase of $t_{\rm age}=0.3~\rm Myr$, the shapes of the profiles in these three runs are very similar to each other, and the peak positions of the absorption feature are almost identical. 
However, at later phases, the shape of the radial profile strongly depends on the stellar mass; 
$t_{\rm decay}$ becomes earlier as the stellar mass increases. 
Fig.~\ref{fig5} shows the time evolution of the escape fractions in these three runs. 
The result with a luminous massive star suggests that the short $t_{\rm decay}$ is originated in the leakage of ionizing photons due to rapid expansion of the ionized region.
In the case of a less massive star,
ionizing photons cannot escape from the halo during the lifetime of the central star so that the deep absorption feature lasts longer as shown by the green dotted curve in Fig.~\ref{fig4}. 
Thus, we find that $\delta T_\mathrm{b}(r)$ sensitively depends on the mass of the first stars. 
We would like to emphasize that such a strong dependence of $\delta T_{\rm b}(r)$ on stellar mass appears only if the absorption of ionizing photons in a mini-halo is appropriately solved. 


\subsection{Halo mass dependence of $\delta T_{\mathrm{b}}(r)$}
\label{sec:mhalo}
\begin{figure}
	\begin{center}
	\includegraphics[bb = 0 0 1296 1296, width=80mm]{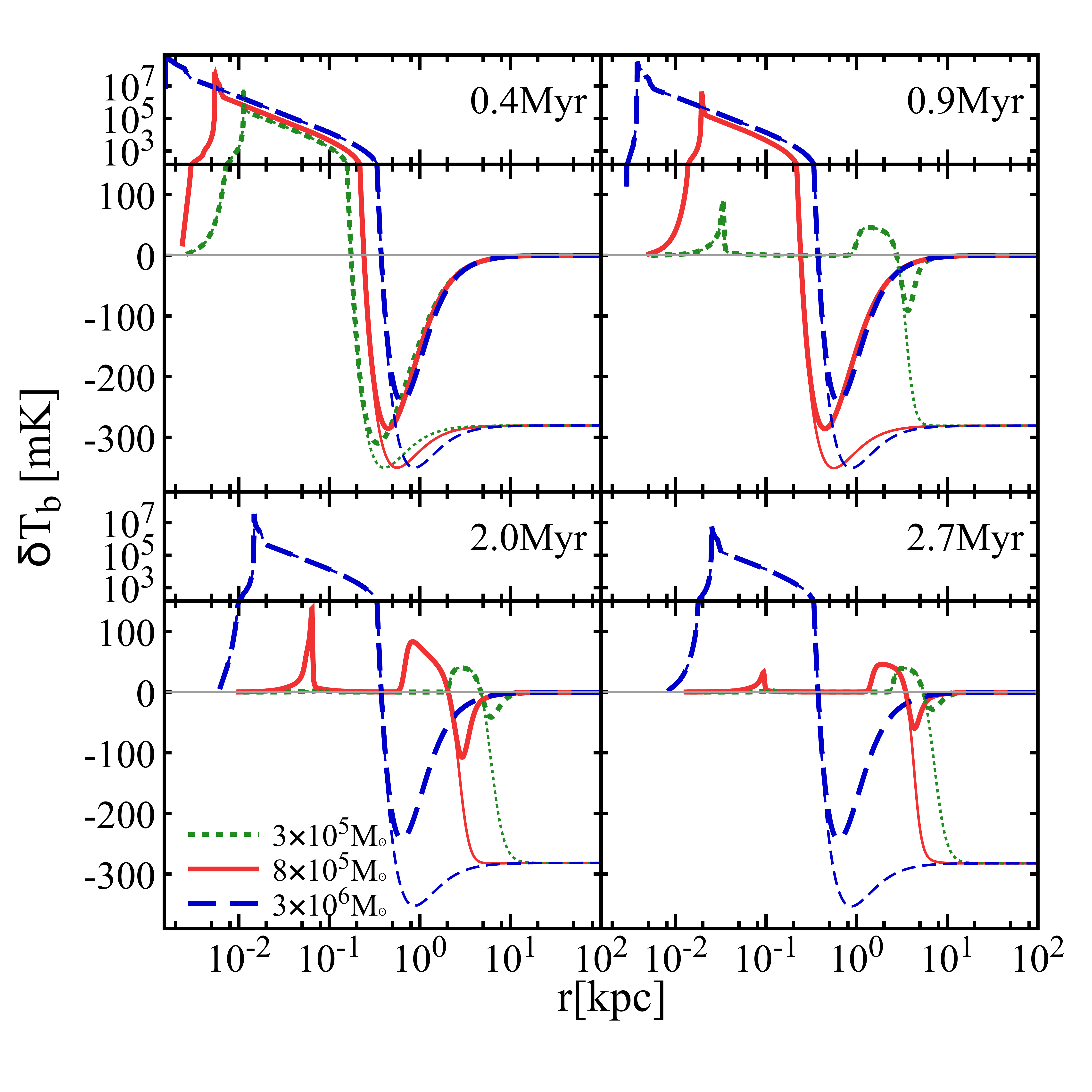}
	\end{center}
	\caption{
	Same as Figure~\ref{fig4}, except for showing Run-z20S100H3e5~(green dotted curve), 
	Run-z20S100H8e5~(red solid curve), and Run-z20S100H3e6~(blue dashed curve). 
	}\label{fig6}
\end{figure}
\begin{figure}
	\begin{center}
	\includegraphics[bb = 0 0 504 504, width=60mm]{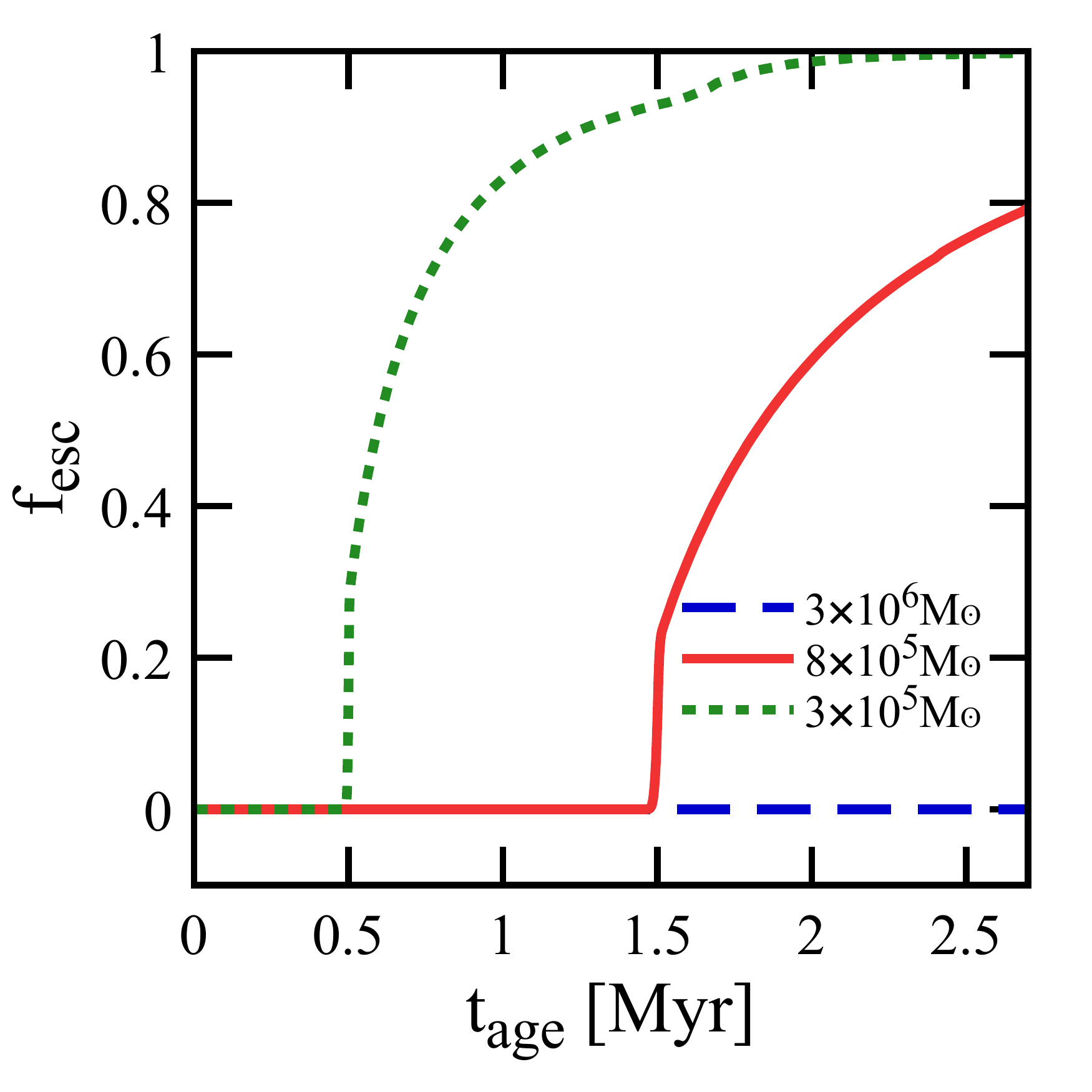}
	\end{center}
	\caption{
	Same as Figure~\ref{fig5}, except for showing Run-z20S100H3e5~(green dotted curve), Run-z20S100H8e5~(red solid curve), and Run-z20S100H3e6~(blue dashed curve).
	}\label{fig7}
\end{figure}
Since the time evolution of $\delta T_{\rm b}(r)$ is sensitive to the dynamics of the ionized gas in a halo as shown in Section~\ref{sec:halogas}, the time evolution is expected to be sensitive to halo mass as well. 
In this section, we perform simulations with various halo masses and examine the halo mass dependence of the 21-cm signal.

Fig.~\ref{fig6} shows the radial profiles of the brightness temperature in Run-z20S100H3e5, Run-z20S100H8e5, and Run-z20S100H3e6, in which different halo masses of $3\times 10^5~\rm M_{\sun}$, $8\times 10^5~\rm M_{\sun}$ and $3\times 10^6~\rm M_{\sun}$ are employed. 

The results at an early phase ($t_{\rm age} = 0.4~\mathrm{Myr}$) show that the absorption feature has its peak position in further distance from the central star for more massive haloes, and also its amplitude is smaller for more massive haloes. 
As discussed in Section~\ref{sec:halogas}, the peak position is determined by the distributions of $n_{\mathrm{H}\textsc{i}}$ and $T_{\rm gas}$. 
More massive haloes are spatially more extended following $r_{\rm vir}\propto M_{\rm halo}^{1/3}$, thus the peak is located farther for more massive haloes. 
This fact also indicates that $\rm Ly\alpha$ flux from the central star at the peak is weaker for more massive haloes, thus its amplitude is smaller for more massive halo. 

In addition, the time evolution from $t_{\rm age} = 0.4$~Myr to $2.7$~Myr shows that $t_{\rm decay}$ becomes longer for more massive haloes.
This delay is caused by the abundant gas and deep gravitational potential of a massive halo. 
Fig.~\ref{fig7} shows the escape fraction of ionizing photons and indeed the rapid increase of the escape fraction is delayed as the halo mass increases. 
As a result, the massive halo with $3\times 10^6~\rm M_{\sun}$ is able to exhibit the deep absorption feature over the lifetime of the central star even if the star is as massive as $100~\rm M_{\sun}$. 

\subsection{Redshift dependence of $\delta T_{\rm b}(r)$}
\label{sec:redshift}
\begin{figure}
	\begin{center}
	\includegraphics[bb = 0 0 1296 1296, width=80mm]{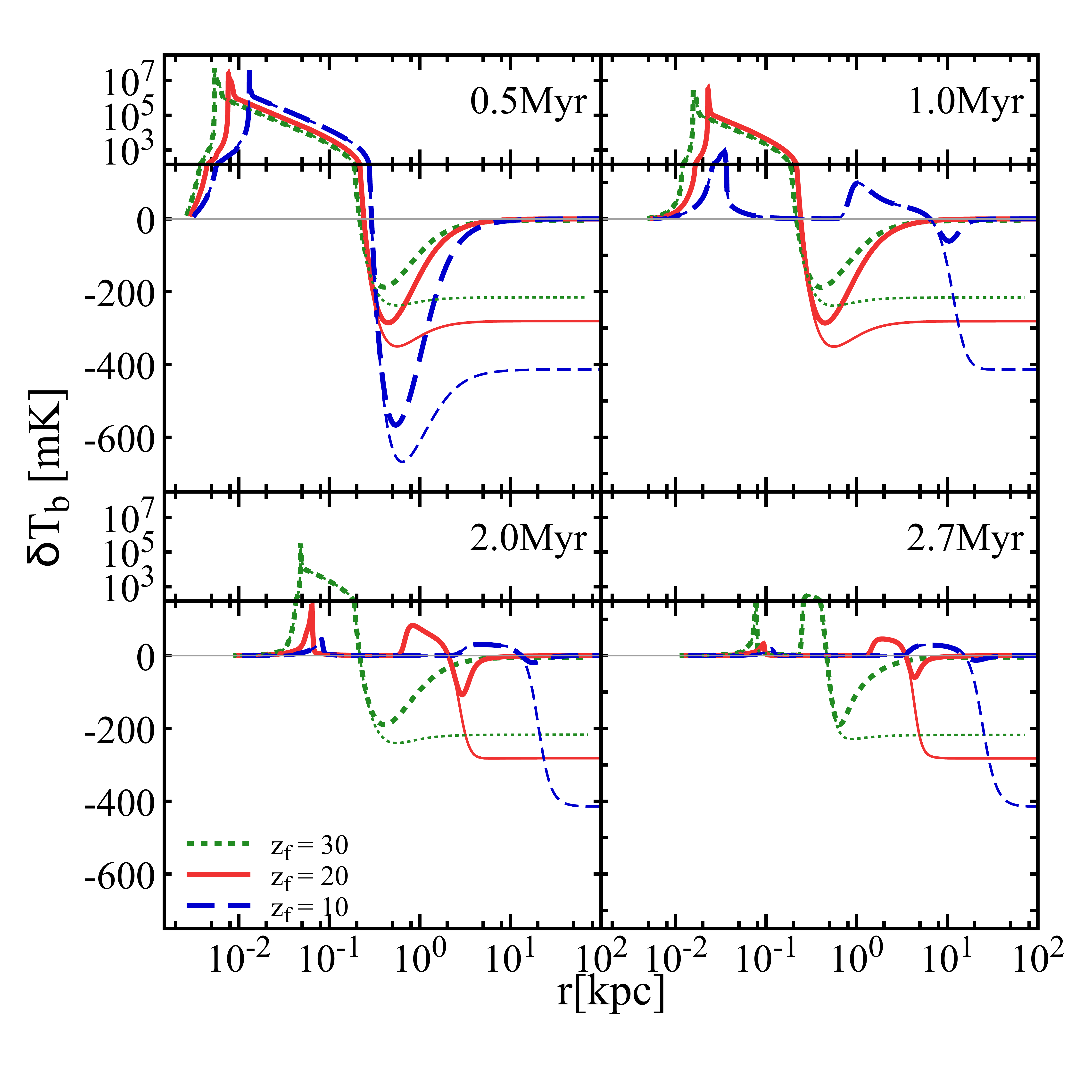}
	\end{center}
	\caption{
	Same as Figure~\ref{fig4}, except for showing Run-z30S100H8e5~(green dotted curve), 
	Run-z20S100H8e5~(red solid curve), and Run-z10S100H8e5~(blue dashed curve). 
	}\label{fig8}
\end{figure}

\begin{figure}
	\begin{center}
	\includegraphics[bb = 0 0 504 504, width=60mm]{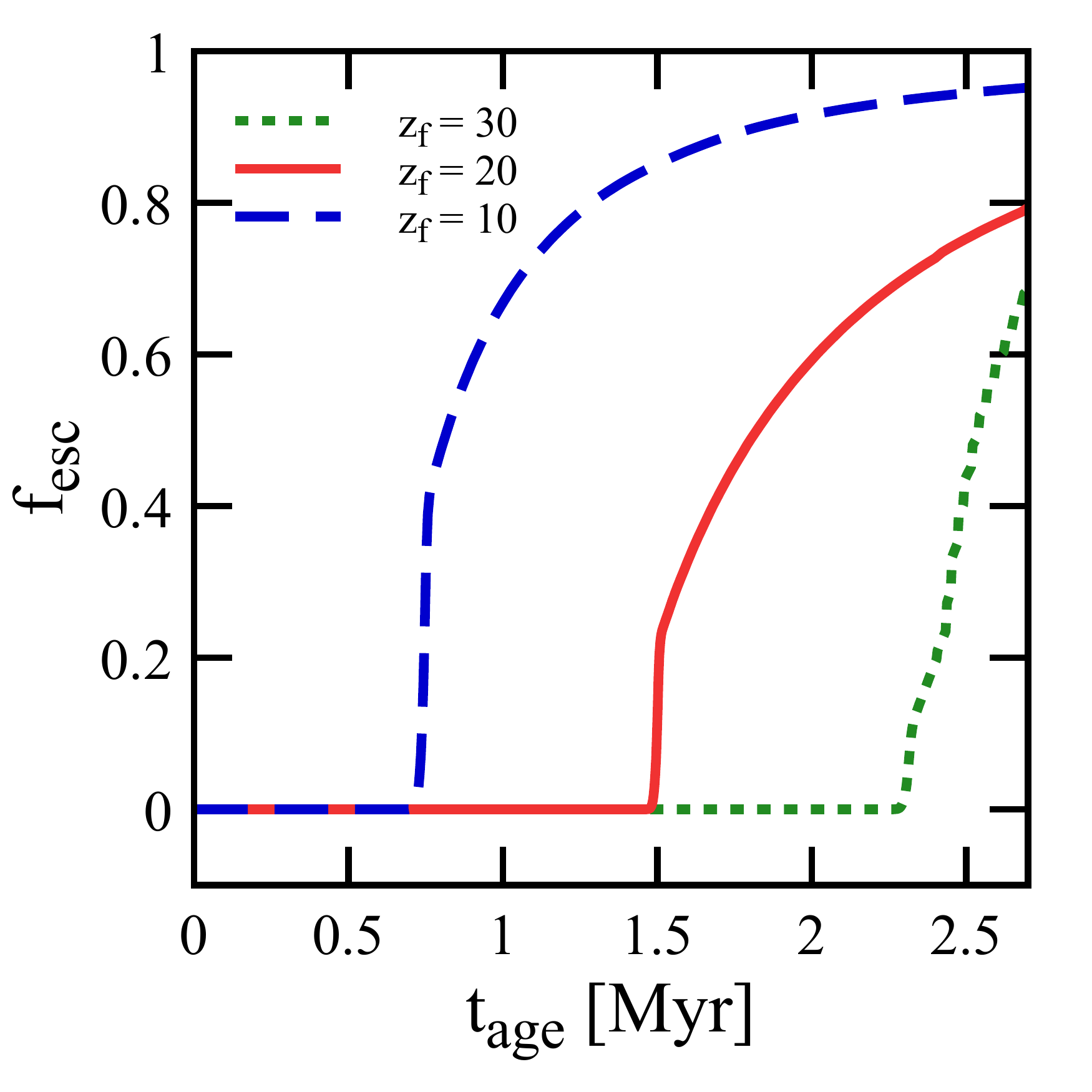}
	\end{center}
	\caption{
	Same as Figure~\ref{fig5}, except for showing Run-z30S100H8e5~(green dotted curve), 
	Run-z20S100H8e5~(red solid curve), and Run-z10S100H8e5~(blue dashed curve).
	 }\label{fig9}
\end{figure}

In this section, we investigate the redshift dependence of the radial profile of the 21-cm signal, which is likely caused by the variation of gravitational potential and the gas density of a halo at different redshifts.

Fig.~\ref{fig8} represents the time evolution of $\delta T_{\rm b}(r)$ in Run-z30S100H8e5, Run-z20S100H8e5, and Run-z10S100H8e5. 
The results at $t_{\rm age}=0.5~\rm Myr$ show that the amplitude of the signal increases as redshift decreases. 
This behaviour is understood as follow; 
the differential brightness temperature is approximately proportional to $-1/\sqrt{1+z_{\rm f}}$, because the gas temperature and the CMB temperature respectively scale as $T_{\rm{gas}} \propto (1+z_{\rm f})^2$ and $T_{\rm{CMB}} \propto (1+z_{\rm f})$. 
In addition, the $\rm Ly\alpha$ coupling coefficient, $x_\alpha$, is proportional to $S_\alpha(z)/T_\mathrm{CMB}(z)$, where $S_\alpha$ weakly depends on redshift. 
These two redshift dependences work so as to make the differential brightness temperature lower for lower redshift. 

As straightforwardly expected, $t_{\rm decay}$ is earlier for lower redshift halo because of the shallower potential, as indicated by the time evolution of the escape fraction in Fig.~\ref{fig9}. 
Therefore, the deep absorption feature around the halo at $z=10$ rapidly disappears, while that at $z=30$ is sustained for long time. 


\subsection{Time evolution of spatially smoothed $\delta T_{\rm b}(r)$}
\label{sec:3.5}
\begin{figure*}
	\begin{center}
	\includegraphics[bb = 0 0 1440 1080, width=130mm]{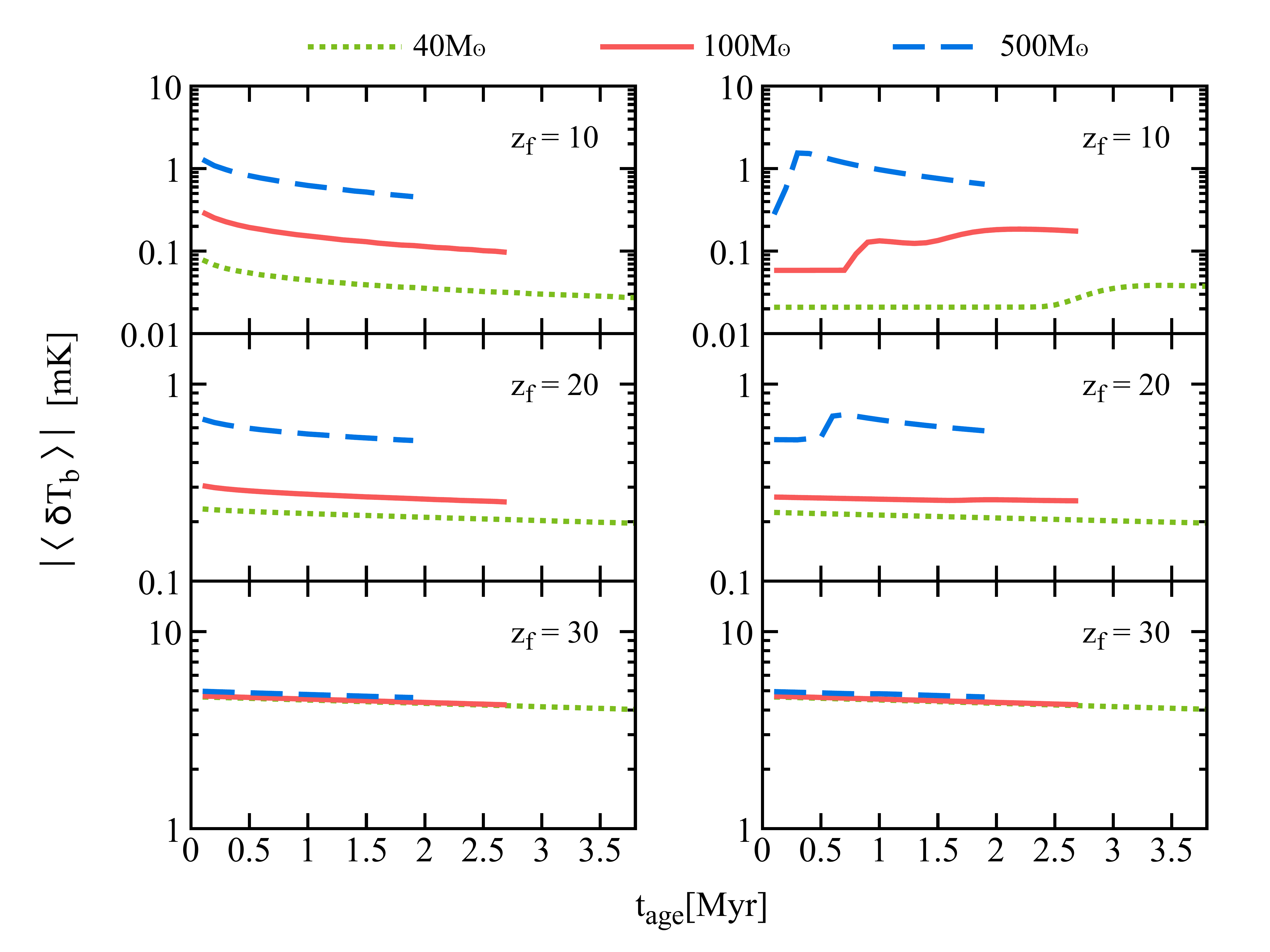}
	\end{center}
	\caption{
	Time evolution of the spatially smoothed differential brightness temperature 
	$ \langle \delta T_\mathrm{b} \rangle$, where the smoothing scale corresponds to 
	$\sim 1~\mathrm{arcmin}$. 
    Note that the absolute values of $\langle \delta T_{\rm b}\rangle$ are shown, though all of the original values are negative. 
	The left and right columns respectively correspond to the cases without and with the dense 
	gas in mini-haloes.  
	In each panel, the time evolution of $ \langle \delta T_\mathrm{b} \rangle$ 
	for $M_\mathrm{star} = 40\mathrm{M}_{\sun}$, 
	$M_\mathrm{star} = 100\mathrm{M}_{\sun}$, and $M_\mathrm{star} = 500\mathrm{M}_{\sun}$ 
	are indicated by the green dotted, red solid, and blue dashed curves, respectively. 
	In the all cases, the halo mass is set to be $M_\mathrm{halo} = 8\times10^5\mathrm{M}_{\sun}$. 
	}\label{fig10}
\end{figure*}
Our results indicate that the radial profile of the differential brightness temperature is sensitive to the properties of the first star.   
Therefore the detection of the 21-cm signal around individual mini-haloes likely provides us with fruitful information. 
Before estimating the detectability of signals from individual stars, which we show in the next section, we here calculate the volume-weighted average of $\delta T_{\rm b}(r)$ (hereafter $ \langle \delta T_\mathrm{b} \rangle$) within an expected angular resolution of the SKA ($1~\mathrm{arcmin}$) and understand the behaviour of the smoothed signal. 

Fig.~\ref{fig10} shows the time evolution of the smoothed signals $\langle \delta T_{\rm b}\rangle$ for various parameter sets. 
Note that we show the absolute values of $\langle \delta T_{\rm b}\rangle$ in Fig.~\ref{fig10}, though all of the original values are negative.  
As shown in previous sections, the deep absorption feature with $\sim-100~\rm mK$ is most notable if the radial profiles of $\delta T_{\rm b}$ can be spatially resolved. 
However, the expected angular resolution of the SKA, which corresponds to $1.5\times10^2$ physical kpc at $z=20$, does not allow us to resolve it. 
Therefore, the very weak 21-cm signals ($\delta T_{\rm b}\sim 0$ at the region extended outer than the absorption region) dominate the area within the scale of the SKA angular resolution and the amplitude of a smoothed signal is only up to
$\sim0.1-1~\rm mK$. 
Besides, the amplitude of a smoothed signal becomes stronger for higher $z_{\rm f}$ despite the fact that the absorption features on spatially resolved profiles show the opposite trend (see Section~3.4).   
The redshift dependence of the smoothed signal is originated from smaller physical scale corresponding to $1~\mathrm{arcmin}$ for higher $z$, i.e. $1.1\times 10^2$ physical kpc at $z=30$ whereas $2.6\times 10^2$ physical kpc at $z=10$. 

The left panels of Fig.~\ref{fig10} show the time evolution of $ \langle \delta T_\mathrm{b} \rangle$ in the cases of a static and uniform medium (e.g. Run-Ref), where the amplitudes of the smoothed signals monotonically decrease with time and increase with stellar mass because the amplitudes are determined by $x_{\rm \alpha}$ which is roughly proportional to $M_{\rm star}/r^2$. 

On the other hand, the time evolution of $ \langle \delta T_\mathrm{b} \rangle$ obtained from our RHD simulations behaves differently as shown in the right panels of Fig.~\ref{fig10}. (To help understanding the behaviour, the blue dashed line in the right middle panel well represents the typical time evolution.)
At the phase of $t_{\rm age}< t_{\rm decay}$, the smoothed signals hardly evolve because the radial profiles are almost steady (see Fig.~\ref{fig1}). 
Although Fig.~\ref{fig1} also shows that the absorption feature on $\delta T_{\rm b}(r)$ starts to decay soon after $t_{\rm decay}$, the rapidly expanding signal region, which works to increase the amplitude, dominantly affects the smoothed signal rather than the decreasing amplitude of the deep absorption, thus the amplitude of $ \langle \delta T_\mathrm{b} \rangle$ rapidly increases at this phase. 
After the sudden increase in $|\langle \delta T_\mathrm{b} \rangle|$, $\rm Ly\alpha$ coupling efficiency decreases due to the dilution of radiation, resulting in the gradually decreasing  $|\langle \delta T_\mathrm{b} \rangle|$. 
Although this sequence is commonly seen in $|\langle \delta T_\mathrm{b} \rangle|$ as far as $t_{\rm decay}< t_{\rm life}$, the earlier phase can only be seen if $t_{\rm decay}$ is late (e.g. Run-z10S100H8e5 indicated by the red solid line in the right-top panel in Fig.~\ref{fig10}).   
In the cases in which $t_{\rm decay}> t_{\rm life}$ is satisfied (e.g., Run-z30S40H8e5 indicated by the green dotted line in the right-bottom panel in Fig.~\ref{fig10}), $ |\langle \delta T_\mathrm{b} \rangle |$ monotonically decreases as the IGM density decreases via the Hubble expansion. 

We emphasize that the amplitudes of the smoothed signals based on the RHD simulations are usually larger than those in the cases with a static and uniform medium after $t_\mathrm{decay}$, because the smoothed signals in the latter cases start to decay soon after the birth of the star (e.g., comparison between the blue dashed curves in the top panels of Figure~\ref{fig10}). 
In contrast, when $t_{\rm age}<t_\mathrm{decay}$, the opposite trend appears (i.e, the amplitudes of the smoothed signals in the former cases are smaller than those in the latter cases), because the signal regions are quite localized within the vicinities of the haloes. 
Thus, roughly speaking, resolving gas in a halo enhances the detectability of its smoothed signal if the first star (the host halo) is massive (less massive) for which $t_\mathrm{decay}$ is typically early. 
\subsection{Detectability of 21-cm signal around a mini-halo}
\label{sec:3.6}
In this section, we compare the smoothed signal $ \langle \delta T_\mathrm{b} \rangle$ with noise temperature expected for the SKA. 
Following \citet{Furlanetto2009}, the noise temperature is approximately given by
 \begin{align}
	\label{skanoise}
	T_\mathrm{noise} \sim 
	1.9\times 10^2 & \left(\frac{10^6~\mathrm{m}^2}{A_\mathrm{eff}} \right)
	\left(\frac{1~\mathrm{arcmin}}{\Delta \theta} \right)^2 \nonumber \\
	& ~~ \times \left(\frac{1+z}{21} \right)^{4.6} 
	\left(\frac{\mathrm{MHz}}{\Delta \nu} 
	\frac{1000~\mathrm{h}}{t_\mathrm{int}} \right)^{0.5}
	~\mathrm{mK}
\end{align}
where $A_\mathrm{eff}$ is the effective collect area, $\Delta\nu$ is the bandwidth,  $\Delta \theta$ is the angular resolution, and $t_\mathrm{int}$ is the integration time. 
We adopt $A_\mathrm{eff} = 10^6$~m$^2$, $\Delta \theta = 1$~arcmin, $\Delta \nu = 1$~MHz, and $t_\mathrm{int} = 1000$~h, respectively. 
According to equation~(\ref{skanoise}), $T_\mathrm{noise} = 9.7$ $\mathrm{mK}$ at $z = 10$, $T_\mathrm{noise} = 1.9\times 10^2$ $\mathrm{mK}$ at $z = 20$, and $T_\mathrm{noise} = 1.2\times 10^3$ $\mathrm{mK}$ at $z = 30$. 

Fig.~\ref{fig10} immediately informs us that the expected noise level is too high to detect the simulated signals around individual haloes, although the time averaged $|\langle \delta T_\mathrm{b} \rangle|$ is enhanced by considering dense gas in a halo and the radiation hydrodynamic effects in some cases. 
One may think that the signals tend to be detectable if we improve the angular resolution. 
However, finer spatial resolutions in observations do not enhance the detectability because the noise level increases faster than the signals.
Therefore we need to increase $A_\mathrm{eff} \times t_{\rm int}^{0.5}$ to detect the individual signals. 
For example, $A_\mathrm{eff}= 2.6\times10^6~\mathrm{m}^2$ and $t_{\rm int}= 1500~\mathrm{h}$ are required for detecting the peak in Run-z10S500H8e5 (the blue dashed curve in the top right panel of Fig.~\ref{fig10}) with $\rm S/N=2$. 

\section{21-cm global signal} \label{sec:global}
In this section, we study the 21-cm global signal which reflects properties of the first stars, such as SFRD ($\dot{\rho}_\star$), and their typical stellar mass. 
The results in Section~\ref{sec:results} show that the radial profiles of the gas temperature and the neutral hydrogen number density are written as functions of $M_{\rm star}$, $M_{\rm halo}$, $t_{\rm age}$, and $z_{\rm f}$ (see Fig.~\ref{fig2}). 
This results allow us to compute the 21-cm global signal based on the following steps for a given parameter set of $M_{\rm star}$, and $\dot{\rho}_\star$. 
The redshift is fixed to be $z=20$ because this is currently the most attractive redshift \citep{Bowman2018}. 

\begin{enumerate}
\item We randomly place 10 stars with $t_{\rm age}$ in a periodic three dimensional calculation box. 
The average volume which a first star occupies equals to $n_\mathrm{star}^{-1}~\mathrm{Mpc}^{3}$, thus the volume of the calculation box containing 10 stars is corresponding to $10n_\mathrm{star}^{-1}~\mathrm{Mpc}^{3}$. Here,  the number density of the first stars, $n_\mathrm{star}$, is given by
\begin{equation}
	n_\mathrm{star} =
	\frac{\dot{\rho}_\star t_\mathrm{life}}{M_\mathrm{star}},
	\label{eq:nstar}
\end{equation}
which enables us to take into account $t_\mathrm{life}$ previous works did not consider. 
The number of grids is determined so as to let the spatial resolution be $4 ~\mathrm{physical~kpc}$~\footnote{We confirmed that estimated global signals hardly change even if we improve the grid size by an order of magnitude. }.
The probability distribution of $t_{\rm age}$ obeys the uniform probability with $0\leq t_{\rm age} \leq t_{\rm life}$. 
Whereas the masses of haloes hosting the stars are randomly determined according to the Press-Schechter mass function raging from $3\times 10^5~\mathrm{M}_{\sun}$ to $4\times 10^6~\mathrm{M}_{\sun}$~\footnote{We divide the halo mass range into 5 bins of $M_\mathrm{halo} = 4\times10^5\mathrm{M}_{\sun}, ~6\times10^5\mathrm{M}_{\sun}, ~8\times10^5\mathrm{M}_{\sun}, ~10^6\mathrm{M}_{\sun}$, and  $2\times10^6\mathrm{M}_{\sun}$, 
for which we additionally performed runs. }. 
The lower limit of the mass function roughly corresponds to the minimum halo mass above which $\rm H_2$ cooling effectively works \citep[e.g.][]{Tegmark1997,Nishi1999}. 
\item We assign $T_{\rm gas}$ and $n_{\mathrm{H}\textsc{i}}$ to all grids, by referring to the radial profiles of $T_{\rm gas}$ and $n_{\mathrm{H}\textsc{i}}$ obtained from the RHD simulations. 
We allow the overlaps of haloes: 
we calculate $n_{\mathrm{H}\textsc{i}}$ by summing up components of the all overlapped haloes, and designate the maximum value of $T_{\rm gas}$ among the overlapped components. 
\item We determine the $\rm Ly\alpha$ coupling coefficients $x_{\alpha}$ at all grids. 
In contrast to the case of an isolated halo shown in Section~\ref{sec:results}, many stars contribute to $x_{\alpha}$ at a given position. 
In this case, redshifted continuum photons mainly contribute to $x_{\alpha}$. 
We extrapolate $x_{\alpha}$ assuming $x_\alpha \propto r^{-2.4}$, and truncate the $x_\alpha$ at the horizon scale of Ly$\beta$ photons (see Section~\ref{calc21cm} for the detailed explanation of the horizon scale). Then, we sum up all values of $x_\alpha$ at each grid to consider the Ly$\alpha$ flux from all stars. 
The scaling relation is motivated by the fact that horizon scales are smaller for higher Lyman series photons \citep{Pritchard2006}. 
\item According to $T_{\rm gas}$, $n_{\mathrm{H}\textsc{i}}$, and $x_{\alpha}$ assigned to all grids, we finally compute $\delta T_{\rm b}$ at the grids with equation~(\ref{eq:dtb}) and (\ref{eq:tspin}), and average them to obtain the global 21-cm signal, $\delta T_\mathrm{b,~global}$. 
Note that the halo components increase the total hydrogen mass in the calculation box, resulting in an artificially high  $\delta T_\mathrm{b,~global}$.
Thus, we rescale $\delta T_\mathrm{b,~global}$ so as to compensate the artificial increase in the total hydrogen mass~\footnote{Without the rescaling, $|\delta T_\mathrm{b,~global}|$ turns out to be significantly higher than $\sim 200~\rm mK$ in the case of $40~M_{\sun}$ and $\dot{\rho}_\star=5\times 10^{-3}~\mathrm{M}_{\sun}~ \mathrm{yr}^{-1}~\mathrm{Mpc}^{-3}$. }. 
\end{enumerate} 

We repeat these steps 2000 times to evaluate the mean value and the variance of the global 21-cm signals for a given parameter set of $M_{\rm star}$ and $\dot{\rho}_\star$. 

\begin{figure}
	\begin{center}
	\includegraphics[bb = 0 0 612 432, width=80mm]{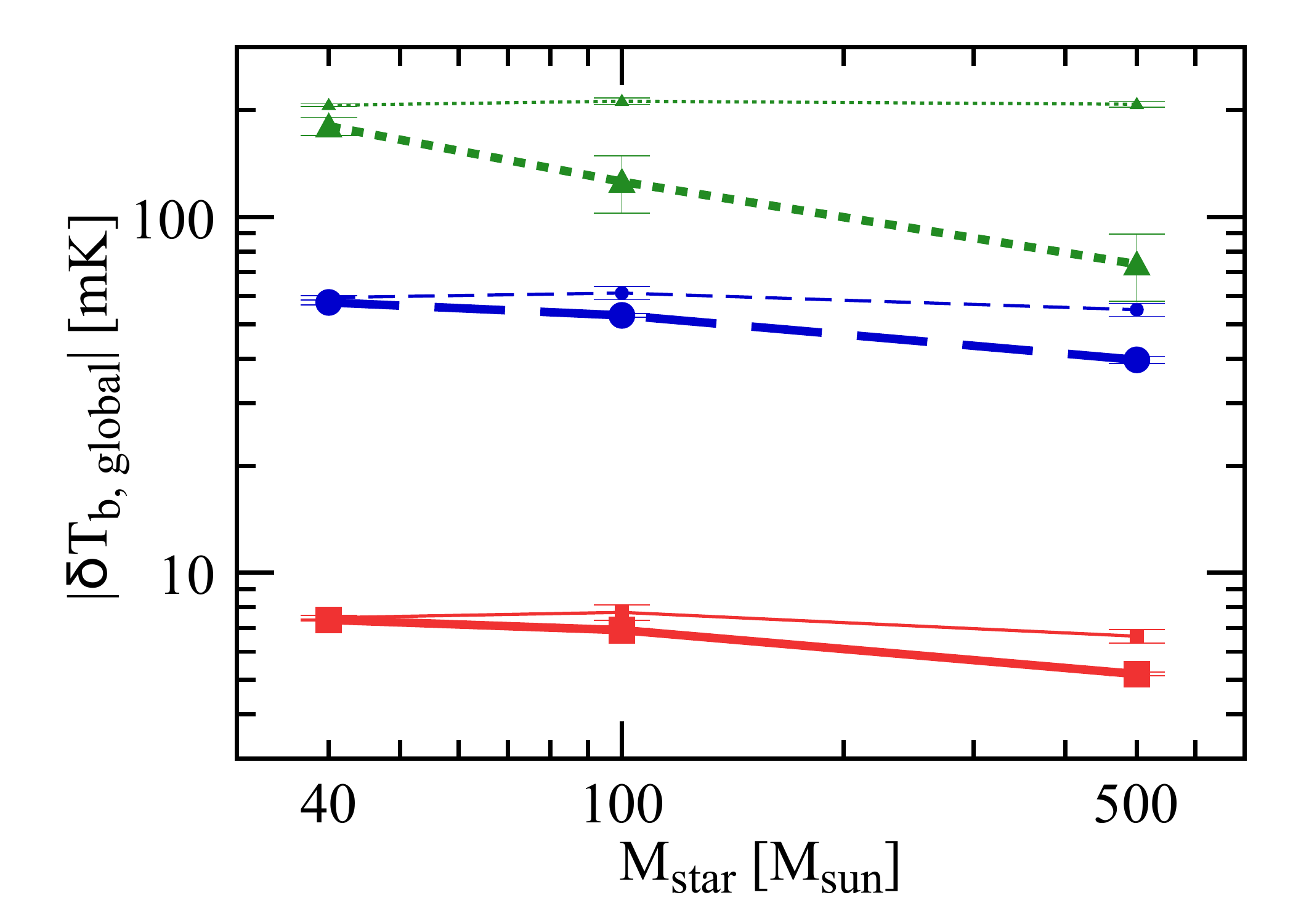}
	\end{center}
	\caption{
	Absolute value of global 21-cm signals at $z = 20$ for $\dot{\rho}_\star = 5\times10^{-5}$ $\mathrm{M}_{\sun}~\mathrm{yr}^{-1}~\mathrm{Mpc}^{-3}$ (red solid line), $\dot{\rho}_\star = 5\times10^{-4}$ $\mathrm{M}_{\sun}~\mathrm{yr}^{-1}~\mathrm{Mpc}^{-3}$ (blue dashed line), $\dot{\rho}_\star = 5\times10^{-3}$ $\mathrm{M}_{\sun}~\mathrm{yr}^{-1}~\mathrm{Mpc}^{-3}$ (green dotted line) as a function of the stellar mass. 
    Note that all of the original values of the global signals shown in this figure are negative. 
	As for the thick lines the temperature distribution is computed based on the results on RHD simulations, 
	whereas $T_{\rm gas}=T_{\rm IGM}$ is assume for the thin lines. 
	}\label{fig11}
\end{figure}

As stated in Section~\ref{intro}, the stellar-mass-dependent lifetime and heating rate, which previous studies did not consider appropriately, are expected to leave their footprints on the global 21-cm signal. 
Fig.~\ref{fig11} shows the obtained global signals for $40~{\rm M_{\sun}} < M_{\rm star} < 500~{\rm M_{\sun}}$ and  $5\times 10^{-5}~{\rm M_{{\sun}}~yr^{-1}~Mpc^{-3}}<\dot{\rho}_\star<5\times 10^{-3}~{\rm M_{{\sun}}~yr^{-1}~Mpc^{-3}}$. 
Let us first discuss how the stellar-mass-dependent lifetime is important, by showing 
the global signals obtained with an assuming of $T_{\rm gas}=T_{\rm IGM}(z)$ (see equation~\ref{eq:tgas}) as the thin lines in Fig.~\ref{fig11}. 
As previous studies \citep{Furlanetto2006, Pritchard2010, Mesinger2013, Yajima2015} showed, the amplitude becomes stronger for higher SFRD, because the WF effect becomes more efficient as the number of stars increases. 
Furthermore, the amplitude decreases with increasing stellar mass: 
roughly speaking, the amount of $\rm Ly\alpha$ photons is proportional to the total stellar mass, 
which follows $\propto \dot{\rho}_\star t_{\rm life}$ (equation~\ref{eq:nstar}). 
As shown by Table~\ref{tb:param}, the lifetime is longer for less massive stars, thus the WF effect effectively works to couple the spin temperature with the gas temperature if the first stars are less massive. 
Note that the stellar mass dependence of the global signal coming from the lifetime is weak. 
For example, with $\dot{\rho}_\star = 5\times10^{-4}$~$\mathrm{M}_{\sun}~\mathrm{yr}^{-1}~\mathrm{Mpc}^{-3}$, the amplitudes are $-59~\rm mK$ for $M_{\rm star}=40\mathrm{M}_{\sun}$,  and $-55~\rm mK$ for $M_{\rm star}=500\mathrm{M}_{\sun}$. 

The discrepancy between the thin and thick lines in Fig.~\ref{fig11} reflects the importance of heating, because the thick lines are obtained by taking photo-heating by the first stars into account. 
In the highest SFRD case of $\dot{\rho}_\star=5\times10^{-3}$~$\rm {M}_{\sun}~{yr}^{-1}~{Mpc}^{-3}$, the global signal is controlled by the average gas temperature owing to the effective WF effect, consequently the discrepancy is most notable in this case. 
We note that the heating is less remarkable if $M_{\rm star}=40~\rm M_{{\sun}}$ even though the total stellar mass is the largest in this case. 
This seemingly strange behaviour is caused by the stellar-mass-dependent escape fraction shown in Section~\ref{sec:results}, where we show $t_{\rm decay}$ is later for less massive stars.  
When $M_{\rm star}=40~\rm M_{{\sun}}$, ionizing photons suffer from the absorption by the gas in haloes, thus hardly heat the IGM even with the highest SFRD. 
In the contrary case of the very massive first stars ($M_{\rm star}=500~\rm M_{{\sun}}$) for which $t_{\rm decay}$ is early, ionizing photons from the first stars well heat the IGM, resulting in the large discrepancy between the thin and thick lines in Fig.~\ref{fig11}. 
We would like to emphasize that it is very difficult to enhance the amplitude of the global signal if the typical mass of the first stars is massive, because the photo-heating rate as well as the $\rm Ly\alpha$ flux increases with SFRD even if we do not consider X-ray radiation. 

To conclude, the stellar mass dependence of the 21-cm global signal especially caused by the time-evolving escape fraction is quite important, though the calculation in this section may be too simplified to quantitatively promise the dependence of the global signal on the SFRD and the first star mass. 

\section{Discussion} \label{sec:discussion}
In this work, we focus only on the 21-cm signature around the main sequence first stars. 
However, in reality even after the lifetime of the central stars, partially ionized regions remain as relic H\textsc{ii} regions. 
\citet{Tokutani2009} found that relic H\textsc{ii} regions are bright in the 21-cm line during the recombination time of the IGM. 
Since the recombination time, which corresponds to an order of 10 Myr at high redshifts, is much longer than the typical lifetime of the first stars, the emission from such relic H\textsc{ii} regions likely contribute to the global 21-cm signal. 
In addition, it has been known that the fate of the first star depends on its mass \citep[e.g.,][]{Heger2002,Umeda2002,Tominaga2007}, and the first star with a certain mass range ends its life as an energetic supernova (SN). 
\citet{Kitayama2005} reported that the SN significantly affects the distribution of the gas temperature and density around a mini-halo. 
A shocks induced by the SN expand outward and sweep up the gas in a relic H\textsc{ii} region, resulting in the enhancement of the detectability of an individual relic region. 
Also high energy photons emitted from the heated gas are expected to affect the IGM temperature and thus impact the 21-cm brightness temperature. 
We further study the stellar mass dependence of the 21-cm signals 
by considering the effects of relic H\textsc{ii} regions, SN, and high-energy photons, in the forthcoming paper. 

\citet{Yajima2014} studied the distinctive 21-cm signatures around the galaxies and quasars, but did not consider the effects by the dense gas in a halo and gas dynamics. 
The effects are also expected to play an important role on the brightness temperature around galaxies and quasars as well as the first stars as shown in the paper. 

When we evaluate the global 21-cm signal, we simply assume that the first stars are randomly distributed. 
However, in reality, the distribution of the stars obeys the background matter distribution, and the number density of the first stars depends on the local density \citep{Ahn2012}. 
Measuring the 21-cm fluctuation arisen by the inhomogeneous distribution of the first stars likely provides us with the information of the clustering of the first stars as well as the typical scale of a 21-cm signal region. 
Hence we are planning to consider such realistic matter distribution in our next study.

\section{Conclusions}\label{sec:conclusion}
In this work, we first performed RHD simulations resolving gas distribution in a mini-halo which previous studies omitted, to investigate the 21-cm signal distribution around the first stars. 
The simulations with such a new point enable us to consider the time-evolving escape fraction. 
When the escape fraction is considerably small, a characteristic deep absorption feature with $\delta T_{\rm b}< -100 ~\mathrm{mK}$ appears at the outer rim of a halo where the gas is colder than the CMB temperature. 
The deep absorption signal starts to decay after the escape fraction rapidly increases at the characteristic time $t_{\rm decay}$, then the signal shape becomes similar to that in previous studies where they assume the static and uniform medium. 
We found for the first time that the resultant radial profile of the brightness temperature strongly depends on the properties of the first stars and the host halo, because the time evolution of the escape fraction is sensitive to the luminosity of the central star and the gravitational potential of the halo as shown by \citet{Kitayama2004}. 
In our simulations, $t_{\rm decay}$ is earlier for more massive stars, for less massive haloes, and for lower redshift haloes, as straightforwardly expected. In some cases that $t_{\rm decay}$ is longer than the stellar lifetime, the absorption feature is sustained until the central star dies. 

Next, we discussed the detectability of the 21-cm signals around individual haloes by comparing the simulated signals with the currently expected specification of the SKA. 
Since an expected angular resolution of the SKA is much larger than the typical virial radius, the radial profile of $\delta T_{\rm b}$ is spatially smoothed out and its amplitude decreases down to $\sim~0.1 - 1~\mathrm{mK}$. 
Thus, the currently expected specification of the SKA is insufficient to detect the individual signal, although the time-evolving brightness temperature in our simulations often enhances the amplitude of the smoothed signal. 
We proposed that $A_\mathrm{eff}\times t_\mathrm{int}^{0.5}  = 1.0\times 10^8~\rm [m^2~h^{0.5}]$ is at least required for the detection of the 21-cm signal around individual haloes. 

We finally investigated how the properties of the first stars are reflected on the 21-cm global signal, by utilizing our simulation results. 
Given SFRD, less massive stars with longer lifetime provide larger amount of $\rm Ly\alpha$ photons than massive stars do, thus the WF effect becomes more effective if the first stars are typically less massive. 
However, this effect turns out to be not notable because of the weak dependence of the lifetime on the stellar mass (see Table~\ref{tb:param}). 
The dominant process controlling the stellar mass dependence of the 21-cm global signal is the heating rate originated in the time-evolving escape fraction described above: the earlier $t_{\rm decay}$ for more massive stars leads to more efficient heating of the IGM. 
An interesting prediction from our results is that the extremely strong absorption feature of the 21-cm global, e.g. $\sim-200~\rm mK$ at $z=20$, is difficult to be reproduced even with an extremely high SFRD if the first stars are typically very massive. 
In summary, to interpret the 21-cm global signals, it is desirable to take into account heating by the first stars, which  strongly depends on $t_\mathrm{decay}$ (thus on the stellar mass). 

\section*{Acknowledgements}
We would like to thank T.~Kitayama for providing us with an RHD code used in this work. All of the simulations were performed with a cluster `galaxy' installed in Nagoya University. 
This work was supported by MEXT's Program for Leading Graduate Schools `Ph.D. Professionals, Gateway to Success in Frontier Asia' (TT), by JSPS KAKENHI Grant Numbers JP17H01110 (KH, NS), JP18K03699 (KH), JP17H04827 (HY), JP18H04570 (HY), 15H05890 (NS), and by Grant-in-Aid for JSPS Research Fellow JP15J04974 and JP18J00508 (MINK).




\bibliographystyle{mnras}
\bibliography{bib} 








\bsp	
\label{lastpage}
\end{document}